\documentclass[preprint,showpacs,showkeys,preprintnumbers,amsmath,amssymb]{revtex4}
\usepackage{dcolumn}
\usepackage{bm}
\usepackage{graphics} 
\renewcommand{\a}{{\alpha}}
\renewcommand{\b}{{\beta}}
\newcommand{\ga}{{\gamma}}
\newcommand{\Ga}{{\Gamma}}
\newcommand{\dl}{{\delta}}
\newcommand{\Dl}{{\Delta}}
\newcommand{\eps}{{\epsilon}}

\newcommand{\la}{{\lambda}}
\newcommand{\hla}{{\hat{\lambda}}}
\newcommand{\La}{{\Lambda}}
\newcommand{\m}{{\mu}}
\newcommand{\n}{{\nu}}
\newcommand{\pa}{{\partial}}
\newcommand{\na}{{\nabla}}
\newcommand{\s}{{\sigma}}
\renewcommand{\r}{{\rho}}
\renewcommand{\th}{{\theta}}

\newcommand{\om}{{\omega}}
\newcommand{\Om}{{\Omega}}
\newcommand{\hOm}{{\hat{\Omega}}}
\newcommand{\hE}{{\hat{E}}}
\newcommand{\hF}{{\hat{F}}}
\newcommand{\hD}{{\hat{D}}}

\newcommand{\eL}{{\mathcal{L}}}

\newcommand{\cF}{{\mathcal{F}}}
\newcommand{\cP}{{\mathcal{P}}}
\newcommand{\cE}{{\mathcal{E}}}
\newcommand{\he}{{\hat{e}}}
\newcommand{\hcL}{{\hat{\mathcal{L}}}}
\newcommand{\ds}{{\displaystyle}}
\newcommand{\imu}{{\rm i}}
\newcommand{\bx}{\bar{x}}

\numberwithin{equation}{section}

\begin{document}
\preprint{}

\title{Seiberg--Witten maps for $\boldsymbol{SO(1,3)}$ gauge
  invariance and deformations of gravity}

\author{S. Marculescu} 
\affiliation{Fachbereich Physik, Universit\"at Siegen, D-57068 Siegen,
  Germany}  
\author{F. Ruiz Ruiz} 
\affiliation{ Departamento de F\'{\i}sica Te\'orica I, Universidad
Complutense de Madrid, 28040 Madrid, Spain}

\date{\today}

\begin{abstract} 
 A family of diffeomorphism-invariant Seiberg--Witten deformations of gravity
 is constructed. In a first step Seiberg--Witten maps for an $SO(1,3)$ gauge
 symmetry are obtained for constant deformation parameters. This includes maps 
 for the vierbein, the spin connection and the Einstein--Hilbert Lagrangian. 
 In a second step the vierbein postulate is imposed in normal coordinates and 
 the deformation parameters are identified with the components 
 $\theta^{\mu\nu}(x)$ of a covariantly constant bivector. This procedure gives 
 for the classical action a power series in the bivector components which by 
 construction is diffeomorphism-invariant. Explicit contributions up to second 
 order are obtained. For completeness a cosmological constant term is included 
 in the analysis.
  
 Covariant constancy of $\,\theta^{\mu\nu}(x)\,$, together with the field
 equations, imply that, up to second order, only four-dimensional metrics
 which are direct sums of two two-dimensional metrics are admissible, the
 two-dimensional curvatures being expressed in terms of $\theta^{\mu\nu}$.
 These four-dimensional metrics can be viewed as a family of deformed
 emergent gravities.
\end{abstract}


\maketitle

\section{\label{sec:One}Introduction}

 It has been known for a long time now that measuring distance with accuracy
 $a$ causes uncertainty $1/a$ in momentum, which, according to the Einstein
 equations, becomes a source of gravitational field. As $a$ decreases, the 
 gravitational field becomes stronger and thus the spacetime curvature grows 
 larger. For $a$ of order of Schwarzschild radius, the gravitational field is 
 strong enough to produce a black hole. In this case, no more information about 
 position is available and a uncertainty relation for position coordinates is 
 due. It has also been known for a while that position uncertainty relations can 
 be realized in terms of non-commutative position operators, provided locality 
 is assumed~\cite{DFR}. The concurrence of these two arguments has triggered an 
 increasing interest in constructing a theory of gravity that includes 
 non-commutative spacetime deformations. See ref.~\cite{Sz} for a recent 
 review. A first step along this direction is the formulation of an effective 
 theory in which the gravitational field, i.e. the spacetime metric is deformed 
 in a way consistent with the principles of general relativity.

 Several proposals for such an effective theory have been
 made~\cite{recent-past,Calmet} in the recent past, including the so-called
 twist-deformed diffeomorphism models~\cite{second,Wess}.  Even though these
 models preserve what are called twisted diffeomorphisms, they violate
 invariance under conventional diffeomorphisms~\cite{AG}. One would like to
 insist on conventional diffeomorphism invariance, among other reasons, to be
 able to observe physics in an frame-independent way. Since everywhere constant
 tensors clash with invariance under general coordinate transformations, one is
 naturally led to consider position-dependent non-commutativity parameters
 $\th^{\m\n}(x)$.

 In this paper we consider an $x$-dependent deformation bivector $\th^{\m\n}$
 and formulate, using Seiberg--Witten maps~\cite{SW}, a theory of deformed
 gravity enjoying diffeomorphism invariance. Such a choice for $\th^{\m\n}$ is
 also favored by string theory. In fact, in all realizations of non-commutative
 spacetimes in string theory~\cite{Chu-Ho}, the non-commutativity parameters
 form an antisymmetric 2-tensor given in terms of a background 2-form 
 $B_2\neq 0$. Furthermore, the open string metric tensor turns out to be given 
 in terms of $\th^{\m\n}$~\cite{SW,Schomerus}. In what follows, we will use the 
 term non-commutative to denote the deformed theory, a widely extended and 
 commonly accepted abuse of language in the literature.

 Our construction of diffeomorphism-invariant non-commutative (NC) deformations
 of gravity is inspired by the description of general relativity as the theory
 that results from imposing the vierbein postulate on an $SO(1,3)$ gauge
 theory~\cite{Kibble}. It consists of two steps. The first one is the
 construction of a Seiberg--Witten gauge theory for $SO(1,3)$ with constant
 deformation parameters $\th^{mn}$.  This construction is algebraic, in
 the sense that it is provided by the solution to a BRS cohomolgy problem, and
 is metric-independent. In the second step gravity is introduced along the
 following lines:
\begin{itemize}
\vspace{-9pt}
\item[(i)] Take normal coordinates with respect to a point $x^a$ at which the 
 Seiberg--Witten construction has been performed. This means that the 
 Christoffel symbols vanish at $x^a$, but not in a neighborhood $\bar{x}^a$ of 
 it. Solve the vierbein postulate in this coordinate system.
\vspace{-6pt}
\item[(ii)] Identify the deformation parameters $\th^{mn}$ with the components
 at $x^a$ of a bivector.  Note that since $\Ga^a_{bc}(x)=0$, this bivector must 
 be covariantly constant in the neighborhood of $x^a$, i.e.
 $\bar{\na}_{\!r} \bar{\th}^{mn}(\bar{x})=0$. This terminates `covariantization' 
 process in the normal coordinate patch and allows transition to the whole 
 four-dimensional manifold. To emphasize this last step Greek indices replace 
 lower case Latin ones.
\vspace{-9pt}
\end{itemize}
\noindent 
 The result will be a Lagrangian which is a power series in $\th^{\m\n}$, whose
 coefficients are functions of the Riemann tensor and its derivatives, and for
 which diffeomorphism invariance is manifest.  We will find explicit expressions 
 for the classical action up to second order in $\th^{\m\n}$.

 Having a prescription to construct a classical action as a power series in
 $\th^{\m\n}$ is not enough to determine if non-commutativity may act as a
 source of gravity. One must elucidate whether the corresponding field
 equations admit solutions for the gravitational field $g_{\m\n}$ with non-zero
 $\th^{\m\n}$. It is worth mentioning in this regard that all NC gravity models
 based on constant non-commutativity proposed so far~ \cite{second,Wess,Calmet}
 yield a vanishing contribution to the classical action at order one in
 $\theta$.

 Equation $\,\na_{\!\r} \th^{\m\n}=0$ relates the spacetime metric with the
 non-commutativity bivector $\th^{\m\n}$.  As is well-known~\cite{Stephani}, it
 only has two solutions, \emph{pp}-wave metrics with null bivectors and
 $(2+2)$-decomposable metrics with non-null bivectors. As we will see below,
 for \emph{pp}-wave metrics the order-two contribution to the classical action
 identically vanishes, whereas for $(2+2)$-decomposable metrics it takes a very
 simple form in terms of two arbitrary parameters. The arbitrariness of these
 parameters arises from the non-uniqueness of the Seiberg--Witten maps, a fact
 well-known for other gauge groups~\cite{Zu}.  We are thus led to the
 conclusion that the only four-dimensional spacetime metrics consistent with
 covariantly constant deformation bivectors are $(2+2)$-decomposable.  This
 limitation on the class of metrics compatible with the appoach proposed here
 has its origin in that the Seiberg--Witten construction involves constant
 deformation parameters. To include other metrics, the Seiberg--Witten
 construction must be extended to also account for derivatives of the 
 deformation parameters $\th^{\m\n}$. See Section \ref{sec:Seven} for a remark 
 on this.

 We emphasize that our approach uses the Moyal--Groenewold product with
 constant deformation parameters.  Our motivation for this is that we are
 interested in setting a deformation procedure that works at any order in the
 deformation parameters. This requires proving BRS covariance for local
 $SO(1,3)$ transformations, and to do so it is essential to have
 associativity, a property guaranteed by this choice of Moyal--Groenewold
 product.  The generally covariant extension along the lines explained above
 breaks breaks associativity, but preserves BRS covariance since by the time
 this is performed one already has a BRS invariant deformed theory. Note that
 since Moyal--Groenewold products with covariantly constant bivectors
 $\th^{\m\n}$ are not associative~\cite{Rivelles,GGBRR}, such bivectors are
 not of Poisson type.

 The paper is organized as follows. In Section \ref{sec:Two}, using general
 covariance arguments that do not rely on any particular deformation of the
 $SO(1,3)$ gauge symmetry, all invariants up to order two in $\th$ that depend
 polynomially on the Riemann tensor and its covariant derivatives and that may
 contribute to the classical action are constructed for metrics satisfying
 $\,\na_{\!\r} \th^{\m\n}=0$. For \emph{pp}-metrics, all invariants of this
 type vanish. For $(2+2)$-decomposable metrics, the number of such invariants
 is sixteen. Out of these sixteen, only two of them contribute to the
 classical action, the contribution being characterized by two arbitrary
 parameters $a$ and $b$. After referring to sections \ref{sec:Four} to
 \ref{sec:Six} for the proof that out of this sixteen invariants, only two
 contribute to the classical action, Section \ref{sec:Three} takes over and
 presents a detailed discussion of the corresponding equations of motion for
 constant curvature. The solutions give the scalar curvatures of the
 two-dimensional metrics in the $(2+2)$-dimensional metric in terms of the NC
 parameters, thus providing a way to classically generate NC gravity. As
 anticipated, sections \ref{sec:Four} to \ref{sec:Six} contain the
 construction of the classical action that was the starting point for Section
 \ref{sec:Three}. This is based on the formulation of a diffeomorphism
 invariant Seiberg--Witten Lagrangian for an $SO(1,3)$ gauge algebra and
 consists of two parts. In Section \ref{sec:Four}, the Seiberg--Witten
 equations for an $SO(1,3)$ gauge symmetry are formulated and a particular
 solution to all orders in $\th$ is found.  This results in a Lagrangian with
 no relation to the underlying spacetime metric and which is not a scalar
 under general coordinate transformations.  Section \ref{sec:Five} explains
 how to impose the vierbein postulate so as to end up with a diffeomorphism
 invariant Lagrangian. Explicit expressions for first and second order
 contributions in $\th$ are computed also in this section. In Section
 \ref{sec:Six}, we find more general solutions to the Seiberg--Witten
 equations which lead to the action taken as starting point in Section
 \ref{sec:Three}.  Finally, Section \ref{sec:Seven} contains our conclusions.
 We also include three Appendices with technical issues.

\section{\label{sec:Two}General structure of Seiberg--Witten deformations up
  to order two}

 We assume that we have a set of constant NC parameters $\vartheta^{\m\n}$ at a
 point $x^\m$ of spacetime. According to the Equivalence principle, it is
 always possible to choose a locally inertial frame centered at that point.
 Since we are interested in invariance under conventional diffeomorphisms, not
 to be confused with twisted diffeomorphisms, the NC parameters
 $\vartheta^{\m\n}$ must be the components of a bivector $\th^{\m\n}\!$.
 Recalling that every bivector constant in a locally inertial frame is
 covariantly constant, one concludes that
\begin{equation}
  \na_{\!\m}\, \th^{\n\r} = 0\,.
\label{constant}
\end{equation}
 It is known~\cite{Stephani} that the only 4-dimensional spacetimes admitting
 covariantly constant bivectors are either \emph{pp}-wave or 
 $(2+2)$-decomposable, so condition~(\ref{constant}) restricts the allowed
 metrics to
\begin{align*}  
 \textnormal{\emph{pp}:}&\qquad ds^2 = du\,dv
      + H(u,x,y)\,du^2 - dx^2 - dy^2\\[3pt]
 2\!+\!2\!: &\qquad
   ds^2 = h'_{\a'\b'}(x^{\a'}) \, dx^{\a'} dx^{\b'}
      + h''_{\a''\b''}(x^{\a''})\,  dx^{\a''} dx^{\b''}\,,
\end{align*}
 where $H$ is an arbitrary function of its arguments, $h'_{\a'\b'}$ and
 $h_{\a''\b''}$ are 2-dimensional metrics and $\a',\b'=0,1$ and
 $\a'',\b''=2,3$. In the first case, the metric can also be written as
 $\,g_{\m\n}=\eta_{\m\n}+ H\,k_\m\/k_\n$, where $\,k_\m=\pa_\m\/u\,$. The
 bivector $\th^{\m\n}$ is null and has the form 
 $\th^{\m\n}= k^\m\/p^\n-k^\n\/p^\m$, with $p^\m$ such that $\,k\cdot\/p=0\,$ 
 and $\,p\cdot\/p=-1$. In the second case the bivector $\th^{\m\n}$ is not null 
 and hence introduces an NC scale, say $\ell_{\rm NC}$.

 The problem of finding the most general $\th$-deformation of the
 Einstein--Hilbert action to order $N$ in $\th$ can be formulated as that of
 constructing all possible invariants of this order using the metric and the
 bivector $\th^{\m\n}$. Let us examine how many of these invariants there are
 at order one and two for both \emph{pp}-wave and $(2+2)$-decomposable
 spacetimes. We restrict ourselves to invariants with polynomial dependence on
 the Riemann tensor and its derivatives.

 At order one, for dimensional reasons, we can only have one bivector
 $\th^{\m\n}$ and either two Riemann tensors $R_{\m\n\r\s}$ or one Riemann
 tensor and two covariant derivatives $\na_{\!\m}$. It is straightforward to
 check that, independently of metric considerations, all invariants of this
 type are identically zero. Let us move on to second order.

 At order two, we must construct all invariants with two $\th^{\m\n}$ and one
 of the following three contents: (i) three Riemann tensors, (ii) two Riemann
 tensors and two covariant derivatives, or (iii) one Riemann tensor and four
 covariant derivatives. Note that invariants without any Riemann tensor are
 trivially zero, since in that case covariant derivatives may only act on
 $\th^{\m\n}$ and this gives zero. To compute the allowed invariants, we 
 rely on the form of the allowed spacetime geometries. Let us first consider the 
 case of \emph{pp}-waves. The Riemann tensor takes the form 
 \hbox{$\,R_{\m\n\r\s} = -2\,k_{[\m}\pa_{\n]} \,\pa_{[\r} H \, k_{\s]}$}. It 
 follows that
\begin{equation*}
   \th^{\m\n}\, R_{\n\a\b\ga}  =  2\,k^\m\, k_\a \, p^\n\, \pa_\n\,
       \pa_{[\b} H \, k_{\ga]} \neq 0 \,,
\end{equation*}
 which in turn implies
\begin{equation*}
   \th^{\m\n}R_{\m\n\r\s}= \th^{\m\n}\,\th^{\r\s}\,R_{\m\r\a\b}=
      \th^{\m\n} R_{\n\b}=0\,.
\end{equation*}
 It is then easy to convince oneself that all invariants of type (i), (ii)
 and (iii) are trivially zero. In other words, there are no
 diffeomorphism-invariant, second-order in $\th$ deformations of \emph{pp}-wave
 metrics.

 Consider now $(2+2)$-decomposable metrics. In this case,
 condition~(\ref{constant}) reduces to
\begin{equation*}
   \na'_{\!\a'}\,\th'^{\b'\ga''} = \na''_{\!\a''}\,\th''^{\b''\ga''} = 0\,,
\end{equation*}
 whose solutions are
\begin{equation*}
   \th'^{\a'\b'}= \frac{\th'}{\sqrt{-h'}\,}~\eps^{\a'\b'} \qquad
   \th''^{\a''\b''}= \frac{\th''}{\sqrt{h''}\,}~\eps^{\a''\b''}\,,
\end{equation*}
 with $\th'$ and $\th''$ constants. Here $\,\eps^{01} =\eps^{23}=1$ and
$\,h'={\rm det}\,(h'_{\a'\b'})\,$ and $\,h''={\rm det}\,(h''_{\a''\b''})$ . The
 four-dimensional bivector $\th^{\m\n}$ is either spacelike or timelike, so the
 NC scale $\ell_{\rm NC}$ is given by
\begin{equation*}
    \th^{\m\n}\th_{\m\n}= 2\,(\th''^2-\th'^2)=\pm\ell^2_{\rm NC}\,.
\end{equation*}
 The only non-zero components of the Riemann tensor are
\begin{equation*}
    R'_{\a'\b'\ga'\dl'}=h'\,\eps_{\a'\b'}\,\eps_{\ga'\dl'}\, R'\qquad
    R''_{\a''\b''\ga''\dl''}= h''\,\eps_{\a''\b''}\,
          \eps_{\ga''\dl''}\,R''\,,
\end{equation*}
 with $R'$ and $R''$ being the Ricci scalars of the the 2-dimensional metrics
 $h'_{\a'\b'}$ and $h''_{\a''\b''}$. Here we have written explicitly factors
 $h'$ and $h''$ so as to have $\,-\eps_{01} =\eps_{23}=1$. In this case there 
 are eleven different invariants. They read
\begin{align}
   \textnormal{Invariants without $\na$'s:}\quad
      & I_1 =  \th'^2 R'^3 - \th''^2 R''^3 \label{I1} \\
      & I_2 =  (R'+R'')\,(\th'^2 R'^2 - \th''^2 R''^2) \label{I2} \\
      & I_3 =  (R'+ R'')^2\,(\th'^2 R' - \th''^2 R'') \label{I3} \\
      & I_4 = (\th'^2-\th''^2)\,(R'+ R'')^3 \label{I4} \\[6pt]
   \textnormal{Invariants with 2 $\na$'s:} \quad
      & J_1 =  \th'^2 \Dl'R'^2 - \th''^2 \Dl''R''^2 \label{J1} \\
      & J_2 =  (R'+R'')\,(\th'^2 \Dl'R' - \th''^2 \Dl''R'')  \label{J2} \\
      & J_3 =  (\th'^2 -\th''^2)\, (\Dl'R'^2 +  \Dl''R''^2) \label{J3} \\
      & J_4 =  (\th'^2-\th''^2)\, (\Dl'+\Dl'')\,(R'+R'')^2 \label{J4} \\
      & J_5 =  (\th'^2-\th''^2)\,(R'+R'')\,(\Dl'R'+\Dl''R'') \label{J5} \\[6pt]
   \textnormal{Invariants with 4 $\na$'s:} \quad
      & K_1 =  \th'^2 \Dl'^2 R' - \th''^2 \Dl''^2 R'' \label{K1} \\
      & K_2 =  (\th'^2-\th''^2)\,(\Dl'+ \Dl'')\,(\Dl'R'+\Dl''R'')\,,
      \label{K2} 
\end{align}
 where $\,\Dl'=h'^{\a'\b'}\na'_{\!\a'}\na'_{\!\b'}\,$ and similarly for
 $\,\Dl''$. 

 If a cosmological constant term is included in the undeformed action, some
 other invariants are possible. On dimensional reasons, the presence of $\La$
 decreases either the number of Riemann tensors by one or the number of
 covariant derivatives by two. At first order in $\th^{\m\n}$, the only
 invariant that may be constructed is $\th^{\m\n}\,R_{\m\n}$, which is
 identically zero.  At order two we may have either (i) two Riemann tensors
 without covariant derivatives, or (ii) one Riemann tensor and two covariant
 derivatives. For \emph{pp}-wave metrics, it is very easy to check that all
 invariants of these types are identically zero. For $(2+2)$-decomposable
 metrics, the list (\ref{I1})-(\ref{K2}) is enlarged with the invariants 
\begin{align}
\textnormal{Invariants for $\La$-term}:
    ~~ & I_5 = \th'^2 R'^2 - \th''^2 R''^2   \label{I5} \\
    & I_6 = (R'+ R'')\,(\th'^2 R' - \th''^2 R'') \label{I6} \\
    & I_7 = (\th'^2-\th''^2)\,(R'+ R'')^2 \label{I7} \\
    & J_6 =  \th'^2 \Dl'R' - \th''^2 \Dl''R''   \label{J6} \\
    & J_7 = (\th'^2 -\th''^2)\, (\Dl'R' +  \Dl''R'') \,. \label{J7} 
\end{align} 
 We conclude that, for \emph{pp}-wave metrics, there are neither 
 first-order, nor second-order polynomial deformations in $\th$ of the 
 Einstein--Hilbert action.  For $(2+2)$-decomposable metrics, the most general 
 deformed Lagrangian up to second order in $\th$ is an arbitrary linear 
 combination of all invariants in (\ref{I1})-(\ref{K2}) and~(\ref{I5})-(\ref{J7}). 
 This is as far as one can go using general invariance arguments. In sections 
 \ref{sec:Four} to \ref{sec:Six} we use the Seiberg--Witten formalism 
 and the vierbein postulate to construct a diffeomorphism invariant NC 
 deformation of the Einstein--Hilbert action. The method yields for 
 $(2+2)$-decomposable metrics the following deformed action up to order two: 
\begin{align}
   S_{2+2} = \frac{1}{\kappa^2} \int d^2\!x'\> d^2\!x'' \>\sqrt{-h'h''}~
   & \bigg\{ \Big( R' +  R'' - \frac{\La}{2}\Big) 
    \Big[\, 1 - \frac{b}{8}\> 
    \Big({\th'}^2 {R'}^2 - {\th''}^2 {R''}^2\,\Big)\Big] \notag  \\
  & + \frac{a}{8}\> 
    \Big(\,{\th'}^2 {R'}^3 - {\th''}^2 {R''}^3\,\Big)\bigg\} + O(\th^3)\,.
\label{NC-action-general}
\end{align} 
 Here $a$ and $b$ are arbitrary real coefficients, their arbitrariness being due 
 to the fact that the solutions to the Seiberg--Witten equations are not unique.

\section{\label{sec:Three}Field equations for deformed gravity and solutions}

The purpose of this section is to show that the equations of motion for the
model described by the classical action~(\ref{NC-action-general}) have
nontrivial solutions. For this purpose, we restrict ourselves to solutions
with constant curvatures $R'$ and $R''$. The field equations then become
algebraic and have the form
\begin{align}
    - \,R' + \frac{\La}{2}  & = \frac{a-b}{8}~
        \Big({\th'}^2 {R'}^3 + 2\,{\th''}^2 {R''}^3\Big)
       + \frac{b}{8}\> \Big[\, \frac{\La}{2}~{\th'}^2 {R'}^2
             + {\th''}^2\, \Big(\frac{\La}{2} - R'\Big) {R''}^2\, \Big]
     \label{eqmo1} \\[3pt]
     R'' - \frac{\La}{2} & = \frac{a-b}{8}~
        \Big( 2\,{\th'}^2 {R'}^3 +  {\th''}^2 {R''}^3 \Big)
     +  \frac{b}{8}\> \Big[\, {\th'}^2\,\Big(\frac{\La}{2} - R''\Big) {R'}^2
        +  \frac{\La}{2}~{\th''}^2 {R''}^2\,\Big] \,.
     \label{eqmo2}
\end{align}
If $R'$ and $R''$ are not constant, the equations above acquire some extra
terms involving covariant derivatives of $R'$ and $R''$, arising from the
higher order terms in the action~(\ref{NC-action-general}). 

We will exclude from our analysis the cases (i) $a=b=0$, for it corresponds to
no deformations at all, and (ii) $R'=R''=0$, for the only solution is then
$\La=0$ and this corresponds to Minkowski spacetime. From Section
\ref{sec:Two} we know that ${\th'}^2\!\neq{\th''}^2$, so at least one of the
two constants $\th',\,\th''$ must be non-zero. Since the equations of
motion~(\ref{eqmo1}) and (\ref{eqmo2}) remain invariant under the changes
\begin{equation}
   (R'',\th'') \leftrightarrow (R',\th') ~~~~~~ (a,b)\leftrightarrow -(a,b)\,,
\label{changes}
\end{equation}
it is enough to consider $\th''\!\neq 0$. The solutions for $\th'\!\neq 0$ are
obtained from those for $\th''\!\neq 0$ by making the replacements above.
Assuming then $\th''\!\neq 0$, we distinguish two types of solutions:
\begin{align*}
  \textnormal{Type 1}\!: ~ &\th''\!\neq 0\,,~\th'\>\textnormal{arbitrary}\,;~
   R',~R''~\textnormal{constants given in terms of}~a,b,\th',\th'' \\
  \textnormal{Type 2}\!: ~ &\th''\!\neq 0\,,~ \th'\!= 0\,;
  ~R'~\textnormal{arbitrary}\,,
  ~ R''=\La= \pm\,\frac{4}{\th''\sqrt{2a}}\,, ~ b=a>0 \,.
\end{align*}
Had we started with $\th'\neq 0$, instead of $\th''\neq 0$, we would have
obtained:
\begin{align*}
  \textnormal{Type 3}\!: ~ &\th'\!\neq 0\,,~\th''\>\textnormal{arbitrary}\,;~
   R',~R''~\textnormal{constants given in terms of}~a,b,\th',\th'' \\
  \textnormal{Type 4}\!: ~ &\th'\!\neq 0\,,~ \th''\!= 0\,;
  ~R''~\textnormal{arbitrary}\,,
  ~ R'=\La= \pm\,\frac{4}{\th'\sqrt{-2a}}\,, ~ b=a<0 \,.
\end{align*}
In the remainder of this section we examine solutions of types 1 and 3.
Actually it is enough to look at type 1, for type 3 can be obtained through
the substitutions~(\ref{changes}). The equations of motion form a system of
two cubic equations in $R'$ and $R''$ with coefficients depending on
$a,b,\th'$ and $\th''$.  It is convenient to distinguish the three following
cases:

\underline{\emph{Case 1}.} One of the 2-dimensional curvatures $R',R''$
 vanishes. After setting one of them equal to zero, eqs.~(\ref{eqmo1})
 and~(\ref{eqmo2}) reduce to two quadratic equations in the non-vanishing
 curvature that can be easily solved. For example, for $R'=0$, non-trivial
 solutions exist only if $\th''\neq 0$. In this case, the curvature $R''$ is 
\begin{equation}
   R''=  \left\{ \begin{array}{ll}
       {\displaystyle \frac{32-3b\La^2{\th''}^2}{\La{\th''}^2(9a-8b)}}
            & \textnormal{if}~~~9a\neq 8b \\[12pt]
              {\displaystyle \frac{3\La}{2}} & \textnormal{if}~~~9a = 8b
                 \end{array} \right. \,,
\label{intable}
\end{equation}
with the cosmological constant given in terms of $\th''$ by
\begin{equation}
    {\bigg(\frac{\La \th''}{2}\bigg)}^{\!2} = \left\{ \begin{array}{ll}
        {\displaystyle\frac{1}{b^3}}\, \Big[\,36\,ab-27\,a^2-8\,b^2 
                           \pm (9a-8b)\,\sqrt{a\,(9a-8b)}\,\Big] 
            & \textnormal{if}~~~b\neq 0\\[9pt]
       {\displaystyle \frac{32}{27\,a}} & \textnormal{if}~~~b=0 
                                              \end{array} \right. \,.
\label{Lapm}
\end{equation}
 Recall that by assumption both $a$ and $b$ cannot vanish simultaneously. The
 right-hand side of eq.~(\ref{Lapm}) must be real and positive. This makes
 clear that not for all $a$ and $b$ a solution $(R''\!,\La)$ exists. Table 1
 collects the allowed ranges for the parameters $a$ and $b$ and the
 corresponding values for $\La^2$. The subscript in $\La^2_\pm$ refers to
\begin{table}[ht]
\begin{center}
\renewcommand\arraystretch{1.2}
\renewcommand\tabcolsep{5pt}
\begin{tabular}{||c|c||} \hline
  \multicolumn{2}{||c||}{Table 1. Solutions with $R'=0$, $\th''\!\neq 0$,}
  \\[2pt]\hline 
 $a<b<0$ &  $\La^2=\La^2_-$ \\\hline
 $b<0,~a\geq 0$ & $\La^2=\La^2_-\,,\La^2_+$ \\\hline
 $b=0,~a>0$ & $\La^2={\displaystyle \frac{128}{27|a|\,{\th''}^2}}$ \\\hline 
 $0<8b\leq 9a$  & $\La^2=\La^2_+$ \\\hline
 $0<8b< 9a\leq 9b$ & $\La^2=\La^2_-\,,\La^2_+$ \\\hline
\end{tabular}
\end{center}
\end{table}
 the $\pm$ sign in front of the square root in ~(\ref{Lapm}). The solutions for 
 $R''=0$ are obtained from those presented here for $R'=0$ through the 
 replacements~(\ref{changes}).

 \underline{\emph{Case 2}.} None of the 2-dimensional curvatures vanishes, but
 the cosmological constant does. For $\La=0$, if $a=b$, the only solution to
 eqs.~(\ref{eqmo1}) and~(\ref{eqmo2}) is the trivial one $R'\!=R''\!=0$.  We
 thus take $a\neq b$. Introducing
\begin{equation*}
   \xi=\frac{R'}{R''} \qquad k=\frac{{\th'}^2}{{\th''}^2}\geq 0\,,
\end{equation*}
 the equations of motion can be written as
\begin{align}
  \frac{8\xi}{{(\th''R'')}^2} & = -\,\big(a-b\big) \big(k\xi^3 +2\big) + b\xi
   \label{red1}\\
  \frac{8}{{(\th''R'')}^2} & = \big(a-b\big) \big(2k\xi^3 +1\big) -bk\xi^2\,.
\label{red2}
\end{align}
 Eliminating $(\th''R'')^2$, one has
\begin{equation}
  k\,\xi^4 + p\, k\,\xi^3 + p\, \xi + 1 =0\,.
\label{master00}
\end{equation}
 where for later convenience we have defined the parameter $p$ as 
\begin{equation*}
    p:=\frac{a-2b}{2\,(a-b)}
\end{equation*}
 Equation~(\ref{master00}) has degree four in $\xi$. Its solutions will depend
 on the parameters $p$ and $k$. We are only interested in real solutions. For
 $p=0$, i.e. for $a=2b$, all solutions are complex. Hence we take $a\neq b,2b$.
 Given a real solution $\xi$, equation~(\ref{red2}) provides $R''$ as a
 function of $a,b,\th$ and $\th''$, and thus a solution $(\xi R'',R'')$ for
 $(R',R'')$.  We must make then sure that equation~(\ref{master00}) has real
 solutions for $\xi$.
 For $k=0$, the only solution to equation~(\ref{master00}) is
\begin{equation*}
  \th'\!=0 \qquad R'\!= - \frac{1}{p}\>R'' 
  \qquad R'' = \pm \frac{2}{\th''}\sqrt{\frac{2}{|a-b|}}\,.
\end{equation*}
 For $k\!>\! 0$, it is shown in Appendix A that equation~(\ref{master00}) has
 real solutions except for
\begin{equation}
    k_+(p^2)\!<k<k_-(p^2)\,,~~ p^2\!<1\,,
\label{restriction}
\end{equation}
 with $k_\pm(p^2)$ given by
\begin{equation} 
  k_\pm(p^2) = \frac{1}{27p^4}\>\Big\{\!\! -  27p^4-2\,(p^2 -4)^3 
        \pm 2\,(p^2 -4)\,(p^2 +8)\,\sqrt{(p^2 -1)\,(p^2 -4)}\,\Big\}\,.
\label{kpm}
\end{equation}
 Hence, for $(\th',\th'')$ with $\th'=\pm\sqrt{k}\th''$ and $k$ as in
 ~(\ref{restriction}) there are no real solutions to the field equations.
 Graphically this is represented in Figure 1, where only the shaded region is
 allowed and the angles $\a_\pm$ are given by $\tan^2\a_\pm=k_\pm$.  
\begin{center}
  \includegraphics{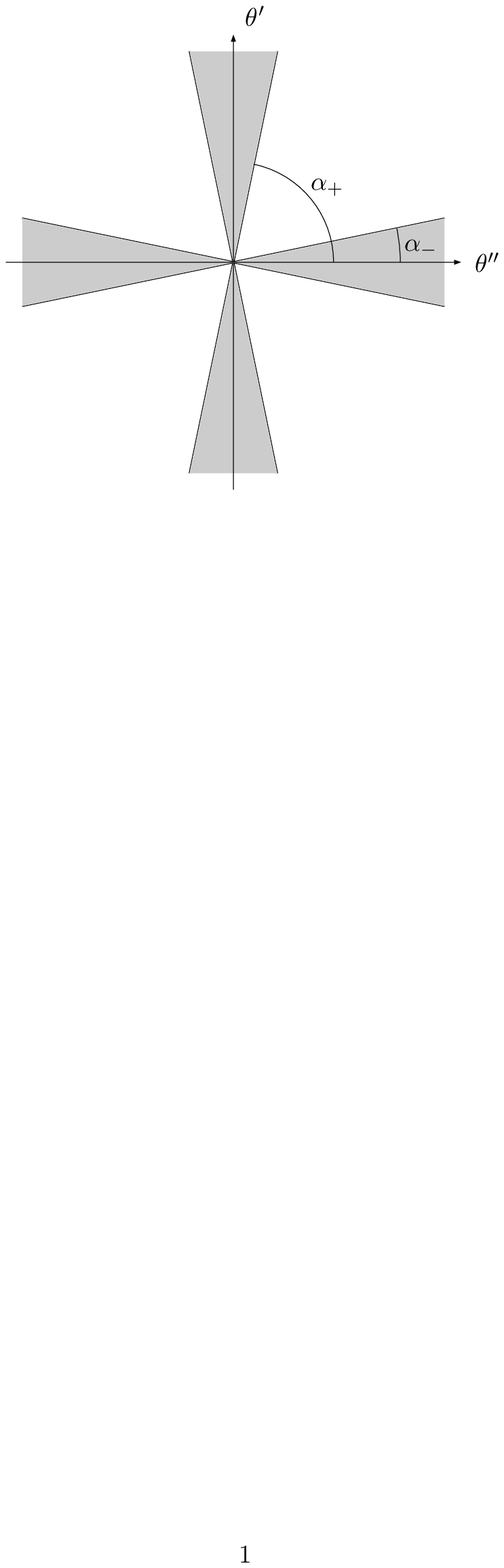}\\
  Figure 1. Allowed region in $(\th',\th'')$-plane for constant curvatures
  $R'$ and $R''$.
\end{center}

 Let us illustrate this case with a simply-looking example. For $p=1$, which
 corresponds to $a=0$, equation~(\ref{master00}) has two real solutions,
 $\xi=-1$ and $\xi= - k^{-1/3}$. The corresponding solutions for $R'$ 
 and $R''$ are
\begin{equation*}
  R'=-R''= \pm \,\frac{4}{\sqrt{\big|b\,\ell^2_{\rm NC}\big|}}
  \qquad \left.\begin{array}{l}
      b<0~~\textnormal{for}~~\th^{\m\n}~~\textnormal{spacelike}\\[3pt]
      b>0~~\textnormal{for}~~\th^{\m\n}~~\textnormal{timelike}
                 \end{array}\right.
\end{equation*}
 and 
\begin{equation*} 
   \th'^{2/3} R' = - \th''^{2/3} R''= \pm\,{\bigg[\,
     \frac{8}{|b\,\big({\th''}^{2/3}-{\th'}^{2/3}\big)|}\,\bigg]}^{1/2}
  \qquad \left.\begin{array}{l}
      b>0~~\textnormal{for}~~\th^{\m\n}~~\textnormal{spacelike}\\[3pt]
      b<0~~\textnormal{for}~~\th^{\m\n}~~\textnormal{timelike}
                 \end{array}\right.
          \,.
\end{equation*}
 The gravitational fields solving the equations of motion can be understood as
 induced by Seiberg--Witten non-commutativity. From this point of view, the
 equations of motion generate a two-parameter family of NC gravities, with
 parameters $a$ and $b$. For instance, the solution above with $R'\!=-R''$
 describes an $\textnormal{AdS}_2\times\textnormal{S}^2$ spacetime, with radii
 proportional to the NC scale,
\begin{equation*}
   R^2_{\rm AdS}=R^2_{\rm S} = \frac{\sqrt{|b|}\,|\ell_{\rm NC}|}{2}\,.
\end{equation*}

\underline{\emph{Case 3}.} None of the curvatures $R',R''$ vanishes, nor the
 cosmological constant. One may proceed as for $\La=0$ and derive an equation
 for $\xi$.  In the general case, that is, leaving aside values of $a$ and $b$
 for which simplifications occur, one obtains an equation of degree nine in
 $\xi$. This guarantees the existence of at least one real solution.  However,
 being an equation of degree nine, an analytic study as that in Appendix A for
 $\La=0$ escapes our abilities.  Yet the equations of motion may be used to
 induce NC gravity, very much as for $\La=0$.  For example, one may be
 interested in 4-dimensional geometries with vanishing curvature, so that
 $R'+R''=0$, but with non-zero cosmological constant. This is achieved e.g. by
 setting $b=0$, for which
\begin{equation*}
    R'=-R''=-\,{\rm sign}\,(a)~
        \frac{3\,|\ell^2_{\rm NC}|\,\La}{8 \,\big({\th'}^2\!+{\th''}^2\big)}
      =-\,{\rm sign}(a)\, \frac{8}{\sqrt{6\,|a\,\ell_{\rm NC}|^2}} \,,
\end{equation*}
 with $a>0$ for spacelike non-commutativity and $a<0$ for timelike. We note
 that even though the action is polynomial in $\th'$ and $\th''$, the
 solutions for the scalar curvatures are not. This makes sense, since one would
 naively expect the NC scale to modify lengths, and dimensional analysis makes
 lengths enter scalar curvatures in a certain way.

 It is worth noting that these solutions do not have a smooth $\th^{\m\n}\to 0$
 limit. This is expected, since $\th^{\m\n}$ must be spacelike and light-like
 deformation bivectors are excluded.  The two-parameter family of
 four-geometries solving the field equations found here can be understood as
 classically induced by the Seiberg--Witten map.  Some solutions for induced or
 emergent NC gravity have been proposed within the context of matrix
 models~\cite{Steinacker}.

\section{\label{sec:Four}Seiberg--Witten maps for $\boldsymbol{SO(1,3)_{\rm
      loc}}$}

 Our construction is based on the description~\cite{Kibble} of general
 relativity as gauge theory with gauge algebra $SO(1,3)_{\rm loc}$. In this
 Section we construct the Seiberg--Witten maps for $SO(1,3)_{\rm loc}$ without
 assuming any metric structure. In the next Section we extend the maps found
 here in a way consistent with diffeomorphism invariance by requiring the
 vierbein postulate.

\subsection{\label{subsec:Four-One}BRS characterization of general relativity}

 We start by reviewing the BRS approach to Kibble's formulation~\cite{Kibble}
 of general relativity as a gauge theory with gauge algebra the local Lorentz
 algebra $SO(1,3)_{\rm loc}$. We will use capital Latin letters $\,A,B,\ldots$
 for $SO(1,3)$ indices. The relevant fields are the vierbein $\,e_\a{}^A(x)\,$
 and the spin-connection $\,\om_\a{}^{AB}(x)$, defined at each spacetime point
 and regarded as independent. 

 The inverse vierbein $\,e_A{}^\a(x)$ is defined as
\begin{equation}
   e_\a{}^A\, e_A{}^\b\!  =\dl_\a{}^\b \qquad
   e_A{}^\a\,e_\a{}^B\! = \dl_A{}^B\,.
\label{inverse}
\end{equation}
 Under a Lorentz transformation, the components of
 $\,e^{A\a}(x)=\eta^{AB}e_B{}^\a(x)\,$ transform for every $\a$ as a vector,
 which we denote by $\,F^\a(x)$. The components of the inverse vierbein
 $e_A{}^\a(x)$ form then its Hermitean conjugate, which we denote by $E^\a$:
\begin{equation*}
    F^\a =\{e^{A\a}\} \qquad
    E^\a = [F^\a ]^+\! = \{e_A{}^\a\}\,\,.
\end{equation*}
 Eqs.~(\ref{inverse}) define the inverse of 
$F^\a(x)$ as $\,\cF_\a(x)=\{e_{\a A}(x) \}$, with
\begin{equation}
   \cF_\a F^\b = \dl_\a^{~\b} \quad F^\a \cF_\a = 1\,.
\label{inverse-F}
\end{equation}
 Taking Hermitean conjugates, the vierbein components form for every $\a$ a
 vector $\,\cE_\a(x)=\{e_\a{}^{A}(x)\}\,$ satisfying
\begin{equation}
   E^\a \cE_\b = \dl^\a_{~\b} \qquad \cE_\a E^\a = 1\,.
\label{inverse-E}
\end{equation}
 The vierbein maps Minkowski's metric $\,\eta_{AB}\,$ onto the object
 $\,\cF_\a(x)\,\cE_\b(x)= e_\a{}^A(x)\,e_\b{}^B(x)\,\eta_{AB}$.  Similarly, the
 inverse vierbein maps $\,\eta^{AB}\,$ onto 
 $\,E^\a(x)\,F^\b(x)= e_A{}^\a(x)\,e_B{}^\b(x)\,\eta^{AB}$.  The transformation 
 properties of $F^\a$ and $\cE_\a$ imply that $E^\a F^\b$ and $\cF_\a\cE_\b$ 
 are invariant under $SO(1,3)$ transformations. Furthermore, since
\begin{equation*}
    \cF_\a \cE_\ga E^\ga F^\b = \dl_\a^{~\b}
    \qquad
     E^\a F^\b \cF_\b \cE_\a = 1\,,
\end{equation*}
 $\cF_\a\cE_\b$ is the inverse of $E^\a\/F^\b$. All these definitions and
 transformation properties are local, i.e. hold at every $x^\a$ independently
 of spacetime metric considerations. We emphasize that $\cF_\a\cE_\b$ is not
 the spacetime metric $g_{\a\b}(x)$, nor $E^\a F^\b$ is its inverse,
 since no assumption relating the spacetime metric and Minkowski's metric
 $\eta_{AB}$ has as yet been made. 

 The spin-connection components $\,\om_\a{}^{AB}(x)\,$ form the matrix
\begin{equation*}
   \Om_\a(x)=- \frac{1}{2}\>\om_\a{}^{AB}(x)\,I_{AB} \,,
\label{VSC}
\end{equation*}
 where we have written $I_{AB}$ for the generators of the vector or adjoint
 representation of $\,SO(1,3)$
\begin{equation*}
   {(I_{AB})}^C{}_D \!=\imu\,\big(\dl_A{}^C\,\eta_{BD}
                     - \dl_B{}^C\,\eta_{AD} \big)\,.
\end{equation*}
 The spin connection is the gauge 1-form for $SO(1,3)_{\rm loc}$, in terms of
 which the Lorentz-covariant derivative $D_\a$ is defined as 
\begin{equation*}
D_\a=\pa_a -\imu\,[\Om_\a,~]\,.
\end{equation*}
 The corresponding field strength $\,\Om_{\a\b}(x)\,$ reads
\begin{equation*}
    \Om_{\a\b}=\pa_\a\Om_\b - \pa_\b\Om_\a - \imu\,[\Om_\a,\Om_\b]\,.
\end{equation*}
 It is trivial to verify that
\begin{equation}
  \big[D_\a,D_\b\big]\, F^\ga = -\imu\,\Om_{\a\b}F^\ga\,.
\label{antisymmetric}
\end{equation}
Furthermore, since $\{F^\a\}$ form a basis, the action of $\Om_{\a\b}$ on
$F^\a$ and its Hermitean conjugate $E^\a$ can always be written as
\begin{equation}
   \imu \,\Om_{\a\b} \,F^\dl = R_{\a \b \ga}{}^\dl \,F^\ga  \qquad
   -\imu\,E^\dl\,\Om_{\a\b} = E^\ga\,R_{\a\b\ga}{}^\dl\, ,
\label{BI}
\end{equation}
where $R_{\a \b \ga}{}^\la$ are real coefficients. In what follows we will
denote by $R_{\a \b}$ the contraction $R_{\a \b}\!:=R_{\a \la\b}{}^\la$. Note
that the coefficients $R_{\a \b}$ are not necessarily symmetric at this stage.
One now considers
\begin{equation}
   \eL= e L \,, 
\label{C-lagrangian}
\end{equation}
with $e$ and $L$ given by
\begin{equation}
   e = \big[- {\rm det}(E^\a F^\b)\big]^{-1/2}
      = \big[- {\rm det}(\cF_\a \cE_\b)\big]^{1/2}\,.
\label{def-e}
\end{equation}
 and
\begin{equation} 
  L = \frac{1}{2\kappa^2}\>\big( \imu E^\a\,\Om_{\a\b}\,F^\b - \La \big) \,.
\label{action-GR}
\end{equation}
It is important to remark that everything so far does not involve any
spacetime metric. We recall in this regard that transformation properties
under diffeomorphisms of tensor fields do not depend on the existence of a
metric. By contrast, partial derivatives of fields do not in general transform
covariantly under diffeomorphisms. In the following we assume that partial
derivatives of fields exist but do not use their transformation laws. In
particular, $\eL$ in (\ref{C-lagrangian}) is not assumed to be a scalar under
general coordinate transformations, for it involves partial derivatives
$\pa_\a\Om_\b$.

To go from $\eL$ and its $SO(1,3)$ local gauge symmetry to general relativity,
one imposes the vierbein postulate
\begin{equation}
   \na_\a \, F^\ga(x) - \imu \, \Om_\a(x) \, F^\ga (x) =0\,,
\label{VP}
\end{equation}
where $\,\na_\a F^\ga(x):=\pa_\a F^\ga(x) + \Ga_{\a \dl}^\ga(x) \, F^\dl(x)\,$
denotes the general covariant derivative. As is well-known, the solution to
eq.~(\ref{VP}), together with the torsion-free assumption
$\,\Ga_{\a\b}^\ga=\Ga_{\b\a}^\ga$, gives the spin-connection $\Om_\a$ in terms
of $F^\a$, its inverse and their Hermitean conjugates.  Furthermore, the
product $E^\a F^\b$ becomes the inverse spacetime metric $g^{\a\b}$ and $\eL$
above the Einstein--Hilbert Lagrangian $\sqrt{-g}(R-\La)/2\kappa^2$, with $R$
the scalar curvature.

The advantage of this approach is that it separates the $SO(1,3)$ gauge
symmetry of general relativity from spacetime metric considerations. This is
very useful to perform Seiberg--Witten deformations of gravity, for the
Seiberg--Witten map provides an algebraic method to construct a deformed
symmetry~\cite{Barnich}. It is most convenient to describe the $SO(1,3)$ local
symmetry in terms of a BRS operator. Let us very briefly recall how this is
done. To remind ourselves that no metric assumptions are made, we will use
lower case Latin letters $\,a,b,\dots$ for indices. Noting that the fields
$F^a,\,\Om_a$ take values in the Lie algebra $SO(1,3)$, we define the BRS
operator $s$ through
\begin{equation}
  s \Om_a = D_a \la
         := \pa_a \la -\imu\,[\Om_a,\la] \qquad
  s F^a = \imu \, \la F^a \qquad
  s \la = \imu \, \la^2 \,,
\label{C-BRS}
\end{equation}
 where $\,\la(x)\,$ is a ghost field taking values in the vector representation
 of $SO(1,3)$. It is straightforward to check that $s^2=0$ and that
\begin{align}
   & s(E^aF^b)=0 \nonumber \\ 
   & s\Om_{ab} =-\imu\,\big[\,\Om_{ab},\la\,\big]\,. \label{C-BRS-Omab} 
\end{align}
 Hence $se=0$.  From this, eqs.~(\ref{C-BRS}) and
\begin{equation*}
   (s X)^+\!=(-)^{|X|}\,sX^+ \qquad |X|=\text{ghost number of}~ X\,,
\label{SCC}
\end{equation*}
 it then trivially follows that $\eL$ is invariant under $s$. The coefficients 
 $R_{abc}{}^d$ introduced in expansions~(\ref{BI}) are also BRS invariant,
\begin{equation*}
    s\big(R_{abd}{}^c\big)=0\,.
\end{equation*}
 Indeed, acting with $s$ on the left-hand side of~(\ref{BI}) and using
 eqs.~(\ref{C-BRS}) and (\ref{C-BRS-Omab}), we have 
 $\,s(\imu\Om_{ab}F^c)\!= R_{abd}{}^c\,(sF^d)$, thus proving BRS invariance of 
 $R_{abd}{}^c$.

\subsection{\label{subsec:Four-Two}Construction of Seiberg--Witten maps}

 Our aim here is to construct an NC extension of the Lagrangian $\eL$ using
 Seiberg--Witten maps. We assume that non-commutativity is characterized by a
 set of BRS invariant constant parameters 
 $\,\th^{ab}\!=\!-\,\th^{ba},\>s\th^{ab}\!  =\!0$, with dimensions of 
 ${\rm (length)}^2$, in terms of which the Moyal--Groenewold 
 \hbox{$\star$-product} reads
\begin{equation}
  \star :=  {\rm exp}\,\big( \frac{\imu\, t}{2}~
     \overleftarrow{\!\pa\!}_{\!a}\,\th^{ab}
     \overrightarrow{\!\pa\!}_{\!b}\big)\,.
\label{MGP}
\end{equation}
The parameter $t$ has been introduced for convenience and takes values on the
interval $[0,1]$. It interpolates between the commutative and the
non-commutative cases.  Throughout this Section $\th^{ab}$ are constant
parameters. As a consequence the Moyal--Groenewold product above is
associative, a property that plays an essential role in performing the
Seiberg--Witten construction to all orders in $\th^{ab}$.

 We recall that, given a set of fields $\{\phi\}$ enjoying a gauge symmetry
 described by a graded, nilpotent BRS operator~$s$
\begin{equation}
    s\phi = \cP(\phi) \,,
\end{equation}
 with $\cP$ a function of $\phi$ and their partial derivatives of finite order,
 the Seiberg--Witten formalism~\cite{SW,Barnich} yields an NC generalization in
 terms of fields $\{\hat{\phi}(t)\}$ defined by
\begin{eqnarray}
  &{\ds s \hat{\phi}(t) = \cP_\star\big[\hat{\phi}(t)\big] } &
  \label{SW-eqn} \\
  &{\ds \hat{\phi}\Big\vert_{t=0}=\phi }
  \label{SW-initial} \,.
\end{eqnarray}
 The functional $\cP_\star$ is obtained from $\cP$ by replacing the ordinary
 product with the star product~$\star$ of eq.~(\ref{MGP}). The fields
 $\{\hat{\phi}(t)\}$ are usually called NC fields, though it is also customary
 to use for them the name of Seiberg--Witten maps.  They are assumed to be
 power series in $\th^{ab}$ whose coefficients are polynomials in the fields
 $\{\phi\}$ and their derivatives, hence taking values in the universal
 enveloping algebra of the gauge Lie algebra under consideration.

 A way to solve equations~(\ref{SW-eqn}), (\ref{SW-initial}) is to differentiate
 the first one with respect to $t$.  This yields
\begin{equation}
   s\,\dot{\hat{\phi}}(t)
        = \frac{d}{dt}\>\cP_\star\big[\hat{\phi}(t)\big]\,.
\label{SW-eqn-dt}
\end{equation}
 The right-hand side is linear in first derivatives $\,\dot{\hat{\phi}}(t)$
 because of the differentiation rule
\begin{equation}
    \frac{d}{dt} \, \Big( \hat{A} \star \hat{B} \Big)
        = \dot{\hat{A}} \star \hat{B}
        + \hat{A} \star \dot{\hat{B}}
        + \frac{\imu}{2} \, \th^{ab} \, \pa_a \hat{A}
           \star  \pa_b \hat{B} \,.
\label{diff-rule}
\end{equation}
 Whether or not the system of equations~(\ref{SW-eqn-dt}) can be solved for
 $\dot{\hat{\phi}}(t)$ must be discussed case by case. Let us assume that a
 solution $\,\dot{\hat{\phi}}_0(t)$ exists.  At $t=0$, it yields the first
 order contribution in $\th^{ab}$, while higher derivatives at $t=0$ provide
 higher contributions. Taking into account the initial condition
 (\ref{SW-initial}), one may then write
\begin{equation*}
   \hat{\phi} = \phi + t\phi^{(1)} + t^2\phi^{(2)} + \ldots\,,
\end{equation*}
 where
\begin{equation*}
   \phi^{(N)} = \frac{1}{N!}~\frac{d^{N-1}}{dt^{N-1}}~
         \dot{\hat{\phi}}(t)\bigg\vert_{t=0}
\end{equation*}
 is the NC contribution to order $N$ in $\th^{ab}$. It is clear that, from this
 point of view, the relevant object to construct the Seiberg--Witten map is
 $\,\dot{\hat{\phi}}_0(t)$.

 Coming to the case we are interested in, the NC fields are now
 $\hat{\phi}=\hOm_m,\,\hF^m,\,\hla$ and take values in the universal enveloping
 algebra of the vector representation of $SO(1,3)$. The Seiberg--Witten
 equations~(\ref{SW-eqn}) read
\begin{equation}
  s \,\hOm_a = \hD_a \star \hla
        := \pa_a \hla - \imu\,[\hOm_a,\hla ]_\star \qquad
  s \,\hF^a = \imu\,\hla\star\hF^a   \qquad
  s \,\hla =  \imu\, \hla \star \hla \,,
\label{NC-BRS}
\end{equation}
 where the notation $\,[X,Y]_\star = X\star Y-Y\star X$ has been used. It is
 straightforward to check that $s^2$ on $\hOm_a,\,\hF^a$ and $\hla$ vanishes. 
 Differentiating equations.~(\ref{NC-BRS}) with respect to $t$, we obtain
\begin{align}
  s\, \dot{\hOm}_a & = \imu \,\big[\hla,\dot{\hOm}_a \big]_\star
                         + \hD_a\star \dot{\hla}
     - \frac{1}{2}~\th^{mn}\, \big\{\pa_m\hla,\pa_n\hOm_a \big\}_\star
  \label{SW-1}\\
  s\, \dot{\hF}^a & = \imu \, \hla\star\dot{\hF}^a
                        + \imu\,\dot{\hla}\star \hF^a
     - \frac{\ds 1}{\ds 2}~\th^{mn}\,\pa_m\hla\star \pa_n\hF^a
  \label{SW-2} \\
  s\, \dot{\hla} & =  \imu\,\big\{\hla,\dot{\hla} \big\}_\star
     - \frac{1}{2}~\th^{mn}\, \pa_m\hla\star \pa_n\hla \,,
  \label{SW-3}
\end{align}
 with $\,\{X,Y\}_\star\! := X\star Y+Y\star X$. One may check by using
 eqs.~(\ref{NC-BRS}) that a particular solution $\,\dot{\hat{\phi}}_0$ is given
 by
\begin{align}
   \dot{\hOm}_a & = \frac{1}{2}~ \th^{mn}\,
       \Big\{ \hOm_m\, , \,- \pa_n\hOm_a
            + \frac{1}{2} \>\hD_a \star \hOm_n\Big\}_\star
   \label{SW-map-1}\\
   \dot{\hF}^a & = \frac{1}{2}~ \th^{mn}\, \hOm_m\star
        \Big( -\, \pa_n \hF^a
        + \frac{\imu}{2}\> \hOm_n \star \hF^a \Big)
   \label{SW-map-2}\\
    \dot{\hla} & = -\,\frac{1}{4}~ \th^{mn} \,
        \big\{ \hOm_m\, , \,\pa_n \hla \big\}_\star \,.
   \label{SW-map-3}
\end{align}

\subsection{\label{subsec:Four-Three}The Seiberg--Witten Lagrangian}

 The NC extension of the Lagrangian~(\ref{C-lagrangian}) is
\begin{equation}
  \hcL = \he\star \hat{L}\,, 
\label{NC-lagrangian}
\end{equation}
 where $\he$ is defined by
\begin{equation}
     \he \star \he \star {\rm det}\,\big(\hE^a\star\hF^b\big) = - 1
\label{NC-det}
\end{equation}
 and $\hat{L}$ reads
\begin{equation} 
    \hat{L}= \frac{1}{2\kappa^2}\> \big( \imu \hE^a\star
          \hOm_{ab}\star \hF^b - \La \big)\,. 
\label{NC-L}
\end{equation}
 Here $\hE^a$ is the Hermitean conjugate of $\hF^a$ and 
 ${\rm det}\,\big(\hE^a\star\hF^b\big)$ is calculated through
\begin{equation*}
   {\rm det}\,\big(\hE^a\star\hF^b\big) = \frac{1}{4 \,!}
       \eps_{a_1a_2a_3a_4}\, \eps_{b_1b_2b_3b_4}\,
           \hE^{a_1}\star\hF^{b_1} \star \hE^{a_2}\star\hF^{b_2}\star
               \hE^{a_3}\star\hF^{b_3} \star \hE^{a_4}\star\hF^{b_4}\,.
\end{equation*}
 Contributions to $\hcL$ of order $N$ in $\th^{ab}$ are given by
\begin{equation}
    \eL^{(N)} = \frac{1}{N!}~\frac{d^{N}}{dt^{N}} ~
           \hcL \bigg\vert_{t=0} \,,
           \label{L-corrections}
\end{equation}
 where derivatives with respect to $t$ are obtained by employing the
 differentiation rule~(\ref{diff-rule}).  Eq.~(\ref{L-corrections}) involves 
 derivatives with respect to $t$ of $\he,\,\hOm_{ab}$ and $\hF^a$. Those of
 $\hOm_{ab}$ and $\hF^a$ follow straightforwardly from eqs.~(\ref{SW-map-1}), 
 (\ref{SW-map-2}), while those of $\he$ are computed by differentiating 
 equation~(\ref{NC-det}) with respect to $t$ as many times as needed. This 
 provides a systematic way to compute the Seiberg--Witten map $\hcL$ to any 
 order in $\th^{ab}$. The algebra may be, and in fact is, long but the method 
 is straightforward. In this paper we consider first and second order 
 corrections in $\th^{ab}$.

 To ease the writing we introduce the notation
\begin{equation*}
    h_{ab}:=\cF_a \cE_b\,,
\end{equation*}
 with $\cF_a$ and $\cE_a$ as in eqs.~(\ref{inverse-F}), (\ref{inverse-E}). For
 the order-one correction we obtain, after some algebra,
\begin{equation}
  \eL^{(1)} = e^{(1)} L + e L^{(1)}
      + \frac{\imu}{2}~\th^{mn}\> (\pa_m e)\, \pa_n L \,,
\label{cL1}
\end{equation}
 where $L$ is as in~(\ref{action-GR}), and $\,e^{(1)}$ and $\,L^{(1)}$ are given
 by
\begin{equation}
   e^{(1)} = -\,\frac{\imu}{4}~e\,\th^{mn}\, h_{ab}\,(D_m E^a)\, D_n F^b
\label{e1}
\end{equation}
 and
\begin{align}
  L^{(1)}  = \frac{\imu}{4\kappa^2}~\th^{mn}\,
    & \Big\{ \big(D_m E^a \big) \Om_{ab} D_n F^b
          + \imu\, E^a \Big( \big\{\Om_{ma}\,, \Om_{nb}\big\}
      - \frac{1}{2} \> \big\{\Om_{ab} \,,\Om_{mn} \big\} \Big) F^b
    \nonumber \\
    & + \pa_m \Big[ \big(D_n E^a\big) \Om_{ab} F^b
                       + {\rm h.c.} \Big] \bigg\} \,.
    \label{L1}
\end{align}
 The second order contribution $\eL^{(2)}$ reads
\begin{align}
   \eL^{(2)} & = e^{(2)} L + e^{(1)} L^{(1)} +  eL^{(2)}
     \notag \\
   &  + \frac{\imu}{2}~\th^{mn}\, \Big[ \,
            \big(\pa_m e^{(1)}\big)\,\pa_n L
          + \big(\pa_m e\big)\,\pa_n L^{(1)}\,\Big]
   - \frac{1}{8}~\th^{mn}\th^{rs}\,\big(\pa_m\pa_r e\big)\,\pa_n\pa_s L\,.
\label{cL2}
\end{align}
 The quantities in this equation have the following expressions: $e^{(2)}$
 is the order-two contribution to $\he$ and has the form
\begin{align}
  e^{(2)} & = e\,\th^{mn} \,\th^{rs}\, \bigg\{ 
    \frac{1}{16}\>h_{ab} \, \big[\, (D_m D_r E^a)\,D_n D_s F^b 
        - 2\imu\, (D_m E^a)\,\Om_{ns}\,D_r F^b\, \big] \nonumber \\
  & -\frac{1}{32}~
    \big( h_{ab}\,h_{cd} + 2\,h_{ad}\,h_{bc} \big)\, 
    \big(D_m E^a\big)\, \big(D_n F^b\big)\, \big(D_r E^c\big) \, D_sF^d 
     \nonumber \\[1.5pt]
  & + \frac{1}{32}~( h_{ab}\,h_{cd} - h_{ad}\,h_{bc} )\, 
      \pa_m  \pa_r  (E^a\/F^b)\,  \pa_n \pa_s (E^c F^d) \nonumber \\[1.5pt]
  & - \frac{1}{12}~\pa_m\pa_r (E^a F^b) \,\Big[
      \big(\pa_n h_{ab} - h_{ab}\,\pa_n \!\ln\! e\big)\> \pa_s\!\ln\! e 
      + \frac{1}{4}\> \big(\, 2 \, h_{bc} \,\pa_n h_{ad} 
      -  h_{ab}\, \pa_n h_{cd}\, \big)\,\pa_s (E^c F^d)\,\Big] 
      \nonumber \\[1.5pt]
  & +\frac{1}{16}\> (\pa_m\pa_r \ln\!e)\, \Big[\, 
      2 \,(\pa_n\!\ln\!e)\,\pa_s\!\ln\!e 
    - 3 \, (\pa_n\pa_s \ln\!e) \Big]\,\bigg\} \,.
\label{e2}
\end{align}
 $L^{(2)}$ is the second order contribution in $\hat{L}$ and can be written as
 the sum
\begin{equation}
   L^{(2)}= L_{\rm v}^{(2)}  +  L_{\rm s}^{(2)}
\label{L2}
\end{equation}
 of two terms given by
\begin{align}
  L_{\rm v}^{(2)} & =  \frac{1}{16\kappa^2}~ \th^{mn}\, \th^{rs} \bigg\{
         \!- \imu  \, (D_m D_r E^a) \, \Om_{ab}\, D_n D_s F^b \notag\\
  & +\Big[\, D_m (E^a\,\Om_{ra}) \, \Om_{sb}\, D_n F^b 
           + D_m (E^a\,\Om_{sb}) \, \Om_{ra}\, D_n F^b \notag \\[1.5pt]
  & \hphantom{+\Big[} 
    - \frac{1}{2}\> D_m (E^a\,\Om_{rs}) \,\Om_{ab} \, D_n F^b 
    + \frac{1}{2}\> D_m (E^a\,\Om_{ab}) \, 
      \big( \Om_{rs}\,D_n F^b - 4\, \Om_{ns}\, D_r F^b \big) 
    + {\rm h. c.} \Big] \notag\\[1.5pt]
  & + (D_m E^a) \, \{\Om_{ab}\,, \Om_{ns}\} \, D_r F^b
    + \frac{1}{2}\> (D_m E^a) \, \big( \{ \Om_{ra}\,, \Om_{sb} \} 
    - 2\, \{ \Om_{ab}\,, \Om_{rs} \} \big) \, D_n F^b  \notag\\
  & + \imu \, E^a \Big[\,2\, \Om_{na}\, \Om_{mr}\, \Om_{sb} 
             - \frac{1}{2}\> \Om_{ma}\, \Om_{rs}\, \Om_{nb} 
             - (a \leftrightarrow b) \Big] \, F^b \notag \\
  & + \imu\, E^a \Big[\, \frac{1}{2}\> \Om_{mn}\, \Om_{ab}\, \Om_{rs} 
             + \big\{ \{ \Om_{na}, \Om_{sb}\} , \Om_{mr} \big\} 
             - \big\{ \{ \Om_{ra}, \Om_{sb}\} , \Om_{mn} \big\} 
       \Big]\, F^b\,\bigg\} 
\label{L2v}
\end{align}
 and 
\begin{align}
  L_{\rm s}^{(2)} & 
      =  \frac{1}{16\kappa^2}~ \th^{mn}\, \th^{rs} \pa_m \bigg\{ 
      \imu\, \big(  D_{(n} D_{s)} E^a\big) \, \Om_{ab} \, D_r F^b 
      \notag\\
      & + \frac{\imu}{2}\> \pa_r \Big[ 2\,(D_n E^a)\, \Om_{ab}\, D_s F^b  
            - \big( D_{(n} D_{s)} E^a \big) \, \Om_{ab} \, F^b \Big]
      \notag \\[3pt]
  & - ( D_r E^a) \Om_{ns} \, \Om_{ab}\, F^b 
    - D_n (E^a\, \Om_{ab})\, \Om_{rs}\, F^b  \notag\\[3pt]
  &  - \frac{1}{2}\> (D_n E^a) \,\Big( 2\,\{ \Om_{ra}\,, \Om_{sb} \} 
        -  \{ \Om_{ab}\,, \Om_{rs} \} \Big)\, F^b \notag \\
  &  - \frac{1}{2}\> \big[ \,D_n (E^a\, \Om_{ra})\, \Om_{sb} 
           + D_n (E^a\, \Om_{sb})\, \Om_{ra}\,\big]  F^b 
     + {\rm h. c.} \bigg\}  \,.
\label{L2s}
\end{align}
 In general, the contribution $\eL^{(N)}$ of order $N$ will be a polynomial of
 degree $N+1$ in $\Om_{ab}$, with covariant derivatives acting on $F^c$ and
 $\,\Om_{ab}F^c$, and on their Hermitean conjugates $E^a$ and $E^c\,\Om_{ab}$.

 There are three important observations concerning the action of partial
 derivatives $\pa_m$ and covariant derivatives $D_m$ in these expressions.  The
 first one concerns $\pa_m$.  From eqs.~(\ref{inverse-F}), (\ref{inverse-E}) it
 is clear that
\begin{equation}
   \pa_m h_{ab} = - h_{ac}\,\pa_m (E^c F^d)\, h_{db} \,.
\label{partial-h}
\end{equation}
 This and the definition~(\ref{def-e}) of $e$ as a linear combination of
 products $E^a F^b$ implies that $\pa_m$ and $\,\pa_m\pa_n$ on $e$ and $h_{ab}$
 are given in terms of $\pa_m (E^a F^b)$ and $\pa_m\pa_n (E^a F^b)$. Noting
 now that 
\begin{eqnarray}
  & \pa_m (E^a F^b) = (D_m E^a)\,F^b + E^a\,D_m F^b 
  & \label{partial-EB} \\
  & \pa_m\pa_n (E^a F^b) = (D_m D_n E^a)\,F^b
      + (D_n E^a)\,D_m F^b + (D_m E^a)\,D_n F^b + E^a\,(D_m D_n F^b), \quad 
    \label{partial2-EB} &
\end{eqnarray}
 we have that $\pa_m$ and $\,\pa_m\pa_n$ acting on $e$ and $h_{ab}$ are linear
 combinations of $D_mE^a,~(D_mE^a)D_nF^b,$ $D_mD_nF^a$ and their Hermitean
 conjugates.

 The second observation concerns the action of products $D_{m}D_{n}$ of two
 covariant derivatives on $E^a$ and $F^b$. Noting eqs.(\ref{antisymmetric}) and
 (\ref{BI}), every product of two covariant derivatives $D_m$ acting on $F^a$
 can then be written as
\begin{equation}
  D_m D_n F^a = -\,\frac{1}{2}~R_{mnb}{}^a F^b
       + \frac{1}{2}~\big\{D_m, D_n \big\}\,F^a\,.
\label{symmetric}
\end{equation} 
We also note here that contributions with $D_m$ acting on
$\,\Om_{ab}F^c=-\imu\, R_{abd}{}^cF^d$, or its Hermitean conjugate
$E^c\,\Om_{ab}$, will yield a term $\,\big(\pa_m R_{abd}{}^c\big)\,F^d$ with
partial derivatives and a term $\,R_{abd}{}^cD_mF^d\,$.
 
 The third observation concerns terms with products of covariant derivatives
 acting on the field strength, $D_{m}\cdots D_{n}\,\Om_{ab}$. In this case,
 integration by parts is performed as many times as necessary until no
 covariant derivatives acting on $\Om_{ab}$ is left and they all act on $F^c$
 and/or $E^c$. This procedure has already been used to obtain
 equations~(\ref{L2v}) and~(\ref{L2v}).

\section{\label{sec:Five}Diffeomorphism invariant Seiberg--Witten Lagrangian}
The Seiberg--Witten Lagrangian $\,\hcL\,$ constructed in the previous section
is a power series in $\th^{ab}$. This construction has been performed in the
universal enveloping algebra of (the vector representation of) $SO(1,3)$ and
is metric-independent.  Note that, precisely because of this metric
independence, covariance under diffeomorphisms only holds for the vierbein and
spin connection, and not for their derivatives.  The problem we face now is to
extend $\,\hcL\,$ to a generally invariant expression without losing BRS
invaraince.

\subsection{\label{subsec:Five-One}The vierbein postulate}
To relate the underlying spacetime metric to the spin-connection $\Om_a$ and
the vierbein $F^a$, we proceed as in general relativity. We take a point $x^a$
to be the origin of a locally inertial frame, impose the vierbein postulate
\begin{equation}
  \big[\bar{\na}_a -\imu\,\Om_a(\bx)\,\big] F^b(\bx) =0
\label{VP-bz}
\end{equation}
 in a neighborhood $\bx^a$ of~$x^a$, and demand a torsion-free geometry
\begin{equation}
  \Ga_{cb}^a (\bx) = \Ga_{bc}^a(\bx) \,.
\label{torsion-free}
\end{equation}
Note that $\bx^a - x^a$ are normal coordinates. The covariant derivative
$\bar{\na}_{\!a}$ at $\bx^a$ in~(\ref{VP-bz}) is defined as usual,
\begin{equation*}
   \bar{\na}_a F^b(\bx) := \bar{\pa}_a \, F^b(\bx)
                         + \Ga^b_{ac}(\bx)\,F^c(\bx)\,,
\end{equation*}
and involves the Christoffel symbols $\Ga^a_{bc}(\bx)$, which depend on the
metric $g_{ab}(\bx)$. Since $\hcL$ is written in terms of fields and their
derivatives at $x^a$, it is convenient to write the vierbein postulate and the
torsion-free condition in terms of fields and derivatives $\na_a$ at $x^a$. In
Appendix B, it is shown that conditions~(\ref{VP-bz}), (\ref{torsion-free})
are equivalent to the infinite set of conditions
\begin{eqnarray}
  &  \big( \na_{a_1} - \imu \, \Om_{a_1} \big) \, \cdots \,
     \big( \na_{a_n} - \imu \,  \Om_{a_n} \big) F^c  ( x ) = 0     &
  \label{SVP} \\[3pt]
  & \Ga_{cd}^b (x) = 0  \qquad\quad
  \pa_{a_1} \, \cdots \, \pa_{a_n} \Ga_{[ c\, d ]}^b ( x ) = 0\,,  &
   \label{STF}
\end{eqnarray}
 with $n = 1, 2, \ldots$ The condition $\Ga^b_{cd}(x)=0$ reminds us that $x^a$
 is the origin of a locally inertial frame. It is convenient to recall that in
 such a frame the derivatives of the Christoffel symbols do not vanish, so that
 conditions~(\ref{STF}) are not trivial.

 The treatment of equations~(\ref{VP-bz}), (\ref{torsion-free}),
 equivalently~(\ref{SVP}), (\ref{STF}), is the same as in general relativity. By
 solving them, the spin-connection $\Om_a(\bx)$ and the Christoffel symbols
 $\,\Ga_{bc}^a(\bx)\,$ are uniquely determined in terms of the (inverse)
 vierbein $\,F^a(\bx)\,$ and its partial derivatives. Once the Christoffel
 symbols are known, it follows trivially that $\,E^a(\bx)F^b(\bx)\,$ is
 covariantly constant and becomes the inverse metric $\,g^{ab}(\bx)$. 
 Furthermore, from equations~(\ref{SVP}), (\ref{STF}) and (\ref{BI}) it follows 
 that
\begin{equation*}
  \big[\na_a-\imu\Om_a,\na_b-\imu\Om_b\big]\, F^c=0 ~ \Rightarrow~ 
  \big[\na_a,\na_b\big]\, F^c=R_{abd}{}^c F^d\,,
\end{equation*}
 so that the coefficients $R_{abd}{}^c$ become the components of the Riemann 
 tensor.

 Having used equations~(\ref{VP-bz}), (\ref{torsion-free}),
 equivalently~(\ref{SVP}), (\ref{STF}), to relate the Lorentz algebra connection
 $\Om_a$ and the vierbein $F^a$ with the underlying metric, we must make sure
 that~(\ref{VP-bz}) and~(\ref{torsion-free}) are compatible with the Lorentz 
 BRS symmetry described by $s$. To  establish the latter, one must show (i) 
 that the spin-connection $\,\Om_a(\bx)$, which is no longer an independent 
 field, transforms under $s$ as in eq.~(\ref{C-BRS}), and (ii) that the 
 Christoffel symbols $\,\Ga_{bc}^a(\bx)\,$ are BRS invariant. Both statements 
 are proved in the Appendix B.

 Although conceptually the situation is similar to general relativity, there is
 a very important technical difference, though. The Seiberg--Witten Lagrangian
 $\hcL$ is far more complicated than the Einstein--Hilbert Lagrangian of general 
 relativity, for it contains products of arbitrary numbers of Lorentz-covariant
 derivatives $\,D_a=\pa_a-\imu\Om_a\,$ acting on the Lorentz field strength 
 $\Om_{ab}$ and the vierbein $F^a$ that must be taken care of. To obtain a
 diffeomorphism-invariant extension of the Seiberg--Witten construct $\hcL$, we
 now proceed in two steps:

\emph{Step~1.} The antisymmetric part of every product of more than two
 covariant derivatives $D_a$ is extracted and all partial derivatives $\pa_a$
 are replaced with covariant derivatives $\na_a$, so that $D_m$ is replaced 
 with $\na_m-\imu\Om_m$. In doing so, Christoffel symbols in $\na_a$ are 
 provided by the solution to the vierbein postulate (\ref{SVP}). A comment 
 concerning the antisymmetrization of products 
 $D_{m_1}\cdots D_{m_n}F^a$ with two or more covariant derivatives is due here. 
 Consider the Lagrangian~(\ref{C-lagrangian}), from which the Einstein--Hilbert 
 action is recovered, and view $\Om_{ab}$ as $D_aD_b-D_bD_a$. Replacing
 $\pa_m\to\na_m$ and blindly using the vierbein postulate~(\ref{SVP}) with
 $n=2$ leads to $\eL=0$. In other words, although not explicitly spelt out,
 antisymmetrization is built in the $SO(1,3)$ gauge description of general
 relativity.

\emph{Step~2.} In Step 1, $\hcL$ has been extended to a power series in the
 parameters $\th^{mn}$ with coefficients generally covariant at $x^a$ and in a
 neighborhood $\bar{x}^a$ of it. To achieve a generally covariant $\hcL_{\rm
   cov}$, a prescription to deal with $\th^{mn}$ is necessary. We identify
 $\th^{mn}$ with the components at $x^a$ of a bivector.  Since this bivector
 has constant components at the origin of a locally inertial frame
 $\,\{\bar{x}^a-x^a\}$, it must be a covariantly constant, i.e. its components
 ${\bar{\th}}^{mn}$ at $\bar{x}^a$ must satisfy $\bar{\na}_{\!r}
 \bar{\th}^{mn}(\bar{x})=0$. This yields an NC Lagrangian $\hcL$ that is
 generally covariant in a normal coordinate patch. Transition to the whole
 four-dimensional manifold is performed in the standard way and Greek indices
 may be restored. In particular the components $\th^{\m\n}$ of the covariantly
 constant bivector satisfy $\na_{\!\r}\th^{\m\n}=0$.

 Following these two steps, the generally invariant extensions of $\eL^{(1)}$
 and $\eL^{(2)}$ will be computed in the next subsection.  As a general word
 of caution, it is convenient to put first the various terms of the
 Seiberg--Witten construct $\hcL$ in a \emph{manifestly} gauge invariant form
 and then apply Steps 1 and 2. 

 We close this Subsection by noting that once Steps 1 and 2 have been
 performed, the original Seiberg--Witten construction with the
 Moyal--Groenewold star product (\ref{MGP}) {\it cannot be recovered.} Indeed,
 generally covariant derivatives do not commute even at the origin of a
 locally inertial frame. Clearly, the deformation parameters in our approach
 do not form a Poisson tensor.

\subsection{\label{subsec:Five-Two}The deformed classical action: explicit
 expressions up to order $\boldsymbol{\th^2}$}

 We first look at $\eL^{(1)}$ in~(\ref{cL1}). From eq.~(\ref{e1}) it follows
 that $e^{(1)}$ is linear in $D_nF^a$ and its Hermitean conjugate $D_nE^a$.
 Step 1 above, i.e. the replacement $D_a\to\na_a-\imu\Om_a$ and the vierbein
 postulate~(\ref{SVP}), then yields that the generally covariant extension of
 $e^{(1)}$ vanishes identically. Similar arguments show that the generally
 covariant extension of $\pa_me$ is also identically zero.  We are thus left
 with the term $eL^{(1)}$ in~(\ref{cL1}) as the only source of generally
 covariant contributions to order one in $\th^{mn}$, the only piece in 
 $L^{(1)}$ that may give a non-trivial contribution being
\begin{equation}
  -\,\frac{1}{4\kappa^2}~\th^{mn}\, E^a 
   \Big( \big\{\Om_{ma}\,, \Om_{nb}\big\}
       - \frac{1}{2} \> \big\{\Om_{ab} \,,\Om_{mn} \big\} \Big) F^b  \,. 
\label{example-L1}
\end{equation}
 Using now eqs.~(\ref{BI}) and recalling from Appendix B that $E^aF^b$ becomes 
 the inverse metric $g^{ab}$ after solving the vierbein postulate~(\ref{SVP}), 
 it is straightforward to see that (\ref{example-L1}) is identically zero. We 
 thus conclude that there is no diffeomorphism-invariant first order deformation 
 of the Einstein--Hilbert action, 
\begin{equation*}
   \eL^{(1)}_{\rm cov}=0\,,
\end{equation*}
 in accordance with the general arguments in Section \ref{sec:Two}.

 Let us next compute the generally invariant extension of the second order
 contribution $\eL^{(2)}$. Inspection of equation~(\ref{cL2}) for $\eL^{(2)}$
 and the arguments used at first order imply that the only non-zero second
 order contributions will arise from terms in $e^{(2)}L$ and $eL^{(2)}$ without
 factors $D_mF^a$ and $D_mE^a$. Form eqs.~(\ref{e2}) and (\ref{L2}) for 
 $e^{(2)}$ and $L^{(2)}$ it follows that there are only three different types
 of such terms:
\begin{itemize}
\vspace{-6pt}
\item[$\bullet$] Terms with products of two or more Lorentz-covariant
 derivatives $D_m$ acting on $F^a$. They are treated as follows. Consider e.g. 
 the first term of $L_{\rm v}^{(2)}$ in~(\ref{L2v}). It gives to $\eL^{(2)}$ a 
 contribution
\begin{equation}
  - \,\frac{\imu\, e}{16 \kappa^2}~ \th^{mn}\th^{rs} \,
       (D_m D_r E^a) \, \Om_{ab}\, (D_n D_s F^b)\,.
\label{example-1}
\end{equation}
 According to Step 1, we use eq.~(\ref{symmetric}) to extract the antisymmetric 
 part of the products $\,D_mD_r$ and $\,D_nD_s$, replace $\pa_n\to\na_{\!n}$ 
 and impose the vierbein postulate~(\ref{SVP}). In accordance with Step 2, we 
 take $\th^{mn}$ as the components at $x^a$ of a generally covariant bivector 
 $\th^{\m\n}$. This gives 
\begin{equation}
       - \,\frac{\sqrt{- g}}{64\kappa^2}~\th^{\m\n}\,\th^{\r\s}\,
       R_{\m\r\ga}{}^\a\,R_{\n\s\dl}{}^\b\, R_{\a\b}{}^{\ga\dl}
\label{example-1-2}
\end{equation}
 for the diffeomorphism invariant extension of~ (\ref{example-1}).
\vspace{-6pt}
\item[$\bullet$] Terms with one covariant derivative $D_m$ acting on  
 $\Om_{ab}F^c$ (or on its Hermitean conjugate $E^c\Om_{ab}$). Consider for 
 example the last term in $L^{(2)}_{\rm s}$, namely
\begin{equation}
  -\, \frac{e}{32\kappa^2}~ \th^{mn}\, \th^{rs} \pa_m \Big[ 
             \,D_n (E^a\, \Om_{ra})\, \Om_{sb}
           + D_n (E^a\, \Om_{sb})\, \Om_{ra}\,\Big]  F^b
     + {\rm h. c.} \, .
\label{example-2}
\end{equation}
 Using expansions~(\ref{BI}), recalling that $D_m$ acts on $R_{abc}{}^d$
 only through $\pa_m$, replacing $\pa_m\to \na_m$, imposing the vierbein
 postulate~(\ref{SVP}) with $n=1$, and noting that after solving the vierbein
 postulate $E^dF^e$ becomes the inverse metric $g^{de}$, one obtains
\begin{equation}
     - \frac{\sqrt{- g}}{16\kappa^2}~ \th^{\m\n}\, \th^{\r\s} \na_\m 
     \Big[ \big(\na_\n R_{\r\dl}\big)\,R_{\s}{}^{\dl} 
            + \big(\na_\n R_{\s\dl\ga\b}\big)\,R_{\r}{}^{\b\ga\dl}\Big]
\label{example-2-2}
\end{equation}
 for the generally covariant extension of~(\ref {example-2}). Here
 $\,R_{\a\b}:=R_{\a\ga\b}{}^{\ga}$ has been used.
 Contribution~(\ref{example-2-2}) is diffeomorphism invariant under the
 assumption $\na_\r\th^{\m\n}=0$.  \vspace{-6pt}
\item[$\bullet$] Terms only involving products of $\Om_{ab},\,E^a$ and $F^a$,
 and no covariant derivatives $D_m$. In this case, all that needs to be used
 are the expansions~(\ref{BI}).  
\vspace{-6pt}
\end{itemize}
 Proceeding in this way with all the terms in $e^{(2)}L$ and $eL^{(2)}\!$, we
 obtain after some algebra the following generally covariant extension for
 $\eL^{(2)}$:
\begin{align}
    \eL^{(2)}_{\rm cov} = \frac{\sqrt{- g}}{16\kappa^2}~& \bigg[\, {\cal R}_1 
     - 2\,{\cal R}_2
        - {\cal R}_3 +\frac{1}{2}\>{\cal R}_4 - \frac{1}{8}\>{\cal R}_5 
        + 2{\cal R}_6 + \frac{1}{4}\>{\cal R}_7 \notag \\ 
   & \! - \frac{1}{2}\>{\cal R}_8 - {\cal R}_{9} + {\cal R}_{10} 
     + \frac{1}{8}\> (R-\La)\,{\cal Q}_1
     +\frac{1}{2}\>{\cal B}_1 - {\cal B}_2\, \bigg] \,, 
   \label{L2-particular}
\end{align}
 the invariants ${\cal R}_1,\ldots,{\cal R}_{10}$ and 
 ${\cal Q}_1,{\cal B}_1, {\cal B}_2$ being 
\begin{alignat*}{3}
    {\cal R}_1 & = \th^{\m\n}\,\th^{\r\s}\, 
                 R_{\m\r}{\!}^{\a\b}\,R_{\n\a\ga\dl}\,R_{\s\b}{\!}^{\ga\dl} 
     & \qquad & {\cal R}_6 = \th^{\m\n}\,\th^{\r\s}\, 
                 R_{\n\a}\,R_{\s\b}\,R_{\m\r}{\!}^{\a\b} \\
    {\cal R}_2 & = \th^{\m\n}\,\th^{\r\s}\, 
                 R_{\m\r}{\!}^{\a\b}\,R_{\n\ga\a}{\!}^\dl\,R_{\s\dl\b}{\!}^\ga
     & & {\cal R}_7 =\th^{\m\n}\,\th^{\r\s}\, 
                 R_{\m\n}{\!}^{\a\b}\,R_{\r\s}{\!}^{\ga\dl}\,R_{\a\b\ga\dl}\\
    {\cal R}_3 & = \th^{\m\n}\,\th^{\r\s}\, 
                 R_{\m\n}{\!}^{\a\b}\,R_{\r\a\ga\dl}\,R_{\s\b}{\!}^{\ga\dl}
     & &  {\cal R}_8 = \th^{\m\n}\,\th^{\r\s}\, 
                R_{\r\a}\,R_{\s\b}\,R_{\m\n}{\!}^{\a\b} \\
    {\cal R}_4 & = \th^{\m\n}\,\th^{\r\s}\, 
                 R_{\m\n}{\!}^{\a\b}\,R_{\r\ga\a}{\!}^\dl\,R_{\s\dl\b}{\!}^\ga
     & &  {\cal R}_{9} = \th^{\m\n}\,\th^{\r\s}\,g^{\ga\dl}\, 
                R_{\n\ga}\,R_{\s\dl\a\b}\,R_{\m\r}{\!}^{\a\b} \\ 
    {\cal R}_5 & = \th^{\m\n}\,\th^{\r\s}\, 
                 R_{\m\r}{\!}^{\a\b}\,R_{\n\s}{\!}^{\ga\dl}\,R_{\a\b\ga\dl}
     & & {\cal R}_{10} =  \th^{\m\n}\,\th^{\r\s}\,g^{\ga\dl}\, 
                R_{\r\ga}\,R_{\s\dl\a\b}\,R_{\m\n}{\!}^{\a\b} 
\end{alignat*}
 and
\begin{align*}
    {\cal Q}_1 & = \th^{\m\n}\,\th^{\r\s}\, R_{\m\r\a\b}\,R_{\n\s}{\!}^{\a\b} \\
    {\cal B}_{1} & = \th^{\m\n}\,\th^{\r\s}\,
          \na_\m\big( R_{\r\b\ga\dl}\,\na_\n\,R_\s{\!}^{\b\ga\dl} \big)\\ 
    {\cal B}_{2} & = \th^{\m\n}\,\th^{\r\s}\,
                   \na_\m\big( R_{\r\b}\,\,\na_\n\,R_\s{\!}^\b \big)\,.
\end{align*} 
 It was already mentioned in Section \ref{sec:Two} that the only spacetime 
 metrics satisfying $\na_{\!\m}\th^{\n\r}=0$ are either of \emph{pp}-wave type 
 or $(2+2)$-decomposable. There it was discussed that all invariants or order 
 two in $\th^{\m\n}$ are identically zero for \emph{pp}-wave metrics. As a
 consistency check one may verify that ${\cal R}_i,{\cal Q}_1$ and ${\cal B}_i$
 vanish for such metrics.  Using the notation of Section \ref{sec:Two} for
 $(2+2)$-decomposable metrics, it is straightforward to check that 
\begin{align}
  4{\cal R}_1 & = - 8{\cal R}_2 = 2{\cal R}_3 = -4{\cal R}_4 
     = 2{\cal R}_5 = 8{\cal R}_6 \nonumber\\
  & = {\cal R}_7 = 4{\cal R}_8 = -4{\cal R}_9  
    = -2{\cal R}_{10} = - 16I_1~~~~~  \label{Rs} \\
  {\cal Q}_1 & = -4\,I_5 \label{Qs}\\[2pt]
  {\cal B}_1 & = {\cal B}_2 = 0 \,,   \label{Bs}
\end{align}
 with $I_1$ and $I_4$ as given in eqs.~(\ref{I1}) and (\ref{I5}). Taking into
 account that $R=2(R'+R'')$ and the expression~(\ref{I5}) for $I_5$, the second
 order Lagrangian becomes 
\begin{equation}
  \eL^{(2)}_{\rm cov, 2+2} = \frac{1}{16\,\kappa^2} \,\sqrt{-h'h''} \, 
     \Big[ I_1 - \Big(R'+ R''-\frac{\La}{2}\,\Big)\,I_5\Big]\,. 
\label{L2-particular-22}
\end{equation}
 This corresponds to taking $a=b=1/2$ in the second order terms of the
 action~(\ref{NC-action-general}).

\section{\label{sec:Six}More general Seiberg--Witten Lagrangians}

 The expression~(\ref{L2-particular}) for the second order contribution
 $\eL^{(2)}_{\rm cov}$ has been found using the Seiberg--Witten
 maps~(\ref{SW-map-1}) and (\ref{SW-map-2}) for $\hOm_a(t)$ and $\hF^a(t)$.
 These are not, however, the most general solutions to the Seiberg--Witten
 equations~(\ref{NC-BRS}), as the following argument shows.  Assume that
 $\,\big\{\hOm_a(t),\,\hF^a(t)\big\}\,$ is a solution to the Seiberg--Witten
 equations~(\ref{NC-BRS}), with $\hla(t)$ as in~(\ref{SW-map-3}). It is then
 clear that $\,\big\{\hOm_a(t)+\dl\hOm_a(t)\,,\,\hF^a(t)+\dl\hF^a(t)\big\}\,$
 is also a solution provided $\dl\hOm_a(t)$ and $\dl\hF^a(t)$
 satisfy~\footnote{In both solutions we have taken the same Seiberg--Witten map
 $\hla$ for the ghost field given by (\ref{SW-map-3}), because it does not
 enter $\hcL$ and, being a ghost field, it is expected to be a gauge-fixing
 artifact.} 
\begin{equation}
   \begin{array}{ll} 
   s\, \dl\hOm_a(t) = \imu\,\big[\hla(t),\dl\hOm_a(t)\,\big]_\star \qquad
    \qquad  & \dl\hOm_a\big|_{t=0}=0\\[6pt]
   s\, \dl\hF^a(t) = \imu\,\hla(t)\star\dl\hF^a(t) \qquad\qquad
      & \dl\hF^a\big|_{t=0}=0\,.
\end{array}
\label{SW-general}
\end{equation}
 To find more general Seiberg--Witten solutions, these two equations must be 
 solved. In what follows we do it. Recall that we are interested in solutions 
 which are formal power series in $\th^{mn}$ whose coefficients depend 
 polynomially in the fields $\Om_a,\,F^a$ and their derivatives.

 Equations~(\ref{SW-general}) are homogeneous in $\dl\hOm_a$ and $\dl\hF^a$,
 and do not contain contributions of order zero in $\th^{mn}$. Their solutions 
 will then be power series in $\th^{mn}$ 
\begin{align*}
   \dl_N\hOm_a & =t^N \big(\dl_N\Om_a\big)^{(N)} 
               + t^{N+1}\,\big(\dl_N\Om_a\big)^{(N+1)} + \ldots \\
   \dl_N\hF^a & = t^N \big(\dl_N F^a\big)^{(N)} 
               + t^{N+1}\,\big(\dl_N F^a\big)^{(N+1)} + \ldots 
\end{align*}
 starting at any order $N\geq 1$ and satisfying
\begin{equation}
   s\,\big(\dl_N\hOm_a\big) = \imu\,\big[\,\hla,\dl_N\hOm_a\,\big]_\star 
   \qquad
   s\, \big(\dl_N\hF^a\big) = \imu\,\hla \star\dl_N\hF^a\,.
\label{SW-general-N}
\end{equation}
 The most general solution for $\dl\hat{\phi}(t)$ will be
\begin{equation*}
   \dl\hat{\phi}(t) = \sum_{N=1}^\infty \dl_N\hat{\phi}(t) 
   \qquad {\phi}=\Om_a,F^a\,.
\end{equation*}
 The form of $\dl_N \hOm_a$ and $\dl_N\hF^a$ can be determined as follows.
 Their lowest-order contributions $\,\big(\dl_N\Om_a\big)^{\!(N)}$ and
 $\,\big(\dl_NF^a\big)^{\!(N)}$ satisfy
\begin{equation}
   s\big(\dl_N\Om_a\big)^{\!(N)} = \imu\,\Big[\,\la\, 
         ,\big(\dl_N\Om_a\big)^{\!(N)}\,\Big]    \qquad
    s\big(\dl_NF^a\big)^{\!(N)} = \imu\,\la\,\big(\dl_NF^a\big)^{\!(N)} 
\label{SW-general-11}
\end{equation}
 for $N=1,2,\ldots$ Using dimensional analysis, BRS covariance and
 eqs.~(\ref{C-BRS}), it is easy to solve these equations. See below for explicit 
 examples. The solutions will be functions $\om^{(N)}_a$ and $\,f^{a(N)}$
\begin{align*}
   \big(\dl_N\Om_a\big)^{\!(N)} & = \om^{(N)}_a\,\big[\Om_m,F^m\big] \,,\\ 
   \big(\dl_NF^a\big)^{\!(N)} & =  f^{a(N)}\,\big[\Om_m,F^m\big]\,
\end{align*}
 of $\Om_m,\,F^m$ and their Lorentz-covariant derivatives. Let multiply 
 $\om^{(N)}_a$ and $\,f^{a(N)}$ with $t^N$ and replace in them the ordinary 
 product with the $\star$-product, the spin connection $\Om_m$ with the full 
 $\hOm_a+\dl\hOm_a$, and vierbein $F^n$ with $\hF^a+\dl\hF^a$. This results 
 in power series
\begin{align*}
   &\, t^N\om^{(N)}_a\,
     \big[\cdot\to\star\,;\,\hOm_m+\dl\hOm_m\,,
                        \,\hF^m+\dl\hF^m\big] \\[1.5pt] 
   & t^N f^{a(N)}\,\big[\cdot\to\star\,;\,\hOm_m+\dl\hOm_m\,,
                        \,\hF^m+\dl\hF^m\big]
\end{align*}
 starting at order $N$ which solve equations~(\ref{SW-general-N}). Hence
\begin{align}
   & \dl_N\hOm_a= t^N \om^{(N)}_a\,\Big[\cdot\to\star\,;\,
      \hOm_m+\sum_{N=1}^\infty \dl_N\hOm_m\,,\,
      \hF^m+\sum_{N=1}^\infty \dl_N\hF^m\Big] \label{gen-sol-Om} \\[1.5pt] 
   & \dl_N\hF^a= t^N f^{a(N)}\,\Big[\cdot\to\star\,;\,
      \hOm_m+\sum_{N=1}^\infty \dl_N\hOm_m\,,\,
      \hF^m+\sum_{N=1}^\infty \dl_N\hF^m\Big]\,. \label{gen-sol-F} 
\end{align}
 provide through iteration explicit solutions to (\ref{SW-general-N}).

 Our interest in this paper are contributions up to second order in $\th^{mn}$.
 It is then enough to consider
\begin{equation}
    \dl\hat{\phi} = t\,\big(\dl_1{\phi}\big)^{\!(1)}
       + t^2\,\big(\dl_1{\phi}\big)^{\!(2)} 
       + t^2\,\big(\dl_2{\phi}\big)^{\!(2)} +\ldots 
\label{enough}
\end{equation}
 for ${\phi}=\Om_a$ and~$F^a$. First-order contributions $(\dl_1{\phi})^{(1)}$ 
 and second-order contributions $(\dl_2{\phi})^{(2)}$ are obtained by solving 
 equations~(\ref{SW-general-11}) for $N=1,2$. In turn, contributions 
 $(\dl_1{\phi})^{(2)}$ are computed by iterating equations~(\ref{gen-sol-Om}), 
 (\ref{gen-sol-F}) once for $N=1$ and by retaining terms quadratic in $t$. All 
 we need are thus the solutions to~(\ref{SW-general-11}) for $N=1,2$. It is 
 straightforward to see that the most general solution to~(\ref{SW-general-11}) 
 for $N=1$ is
\begin{align*}
   \big(\dl_1\Om_a\big)^{\!(1)} & = \om^{(1)}_a 
             = \frac{c}{2}\> \th^{mn}\, D_a\Om_{mn} \\[3pt]
    \big(\dl_1 F^a\big)^{\!(1)} & = f^{a(1)} = \frac{\imu}{2}\,\Big( 
      \,p\,\th^{mn}\,\Om_{mn}F^a + q\,\th^{am}\,\Om_{mb} F^b
      +  r\, \th^{am}\,\big\{ D_m , D_b \big\} F^b \,\Big)\,, 
\end{align*}
 with $c,\,p,\,q$ and $r$ arbitrary real coefficients. Eqs.~(\ref{gen-sol-Om}) 
 and (\ref{gen-sol-F}) for $N=1$ then read
\begin{align}
   \dl_1\hOm_a & = \frac{t}{2}\>c\,\th^{m n}\,\hD_a^{\hOm +\dl\hOm} \star
              \big( \hOm_{mn}+\dl\hOm_{mn}\big)  \label{NCc} \\[3pt]
   \dl_1\hF^a & = \frac{\imu\,t}{2}\> \Big[\> p\,\th^{mn}\, 
      \big(\hOm_{mn} + \dl\hOm_{mn} \big) \star \big( \hF^a+\dl\hF^a\big) 
      \nonumber \\
      & \hphantom{\frac{\imu}{2}~ \Big[\>\,} 
        + q\,\th^{am}\,\big(\hOm_{mb}+\dl\hOm_{mb}\big)\star
               \big( \hF^b+\dl\hF^b\big) \nonumber\\[2pt]
      & \hphantom{\frac{\imu}{2}~ \Big[\>\,}  
        + r\,\th^{am}\,
          \big\{\hD^{\hOm+\dl\hOm}_m , \hD^{\hOm+\dl\hOm}_b\big\}_\star
          \star \big( \hF^b+\dl\hF^b\big) \Big]  \label{NCpq} 
\end{align}
 with $\,\hD_a^{\hOm +\dl\hOm}$ the Lorentz $\star$-covariant derivative
 $\,\hD_a^{\hOm +\dl\hOm} = \pa_a - \imu\,\big[\hOm_a+\dl\hOm_a,~\big]_\star\,$
 and
\begin{equation*}
  \dl \hOm_{mn} = \hD_m \star\dl\hOm_n 
  - \hD_n\star\dl\hOm_m 
  - \imu\,\big[\dl\hOm_m , \dl\hOm_n \big]_\star\,.
\end{equation*}
 From here it follows that
\begin{equation*}
   \big(\dl_1\Om_a\big)^{\!(2)} = \frac{1}{2}\,\frac{d^2}{dt^2}\,
    \Big\{\,\frac{t}{2}\>c\,\th^{mn}\, \Big[\,\hD_a\star\hOm_{mn}
       + t\,D_a\big(\dl_1\Om_{mn}\big)^{\!(1)} 
       - \imu\,t\,\big[\big(\dl_1\Om_a\big)^{\!(1)},\Om_{mn}\big] 
     \Big] \Big\}{\bigg|}_{t=0}
\end{equation*}
 and 
\begin{align*}
   \big(\dl_1F^a\big)^{\!(2)} = \frac{1}{2}\,\frac{d^2}{dt^2}\, 
     &\,\Big\{\, \frac{\imu\,t}{2}\,p\,\th^{mn}\,
     \Big[ \hOm_{mn}\star\hF^a + t\,\Om_{mn}\,\big(\dl_1F^a\big)^{\!(1)} 
         + t\,\big(\dl_1\Om_{mn}\big)^{\!(1)} F^a \Big] \\
  & + \frac{\imu\,t}{2}\,q\,\th^{am}\, \Big[ \hOm_{mb}\star\hF^b 
         + t\,\Om_{mb}\,\big(\dl_1F^b\big)^{\!(1)}
         + t\,\big(\dl_1\Om_{mb}\big)^{\!(1)} F^b \Big] \\
  & + \frac{\imu\,t}{2} \,r\,\th^{am}\, \Big[ \big\{\hD_m , \hD_b\big\}_\star 
      \star\hF^b 
          + t\,\big\{D_m , D_b\big\}\big(\dl_1F^b\big)^{\!(1)} \\
  & \hphantom{\frac{\imu\,t}{2} \,r\,\th^{am}~} 
    - \imu\,t\,\big\{D_m , \big(\dl_1\Om_b\big)^{\!(1)}\big\}F^b
          - \imu\,t\,\big\{D_b , \big(\dl_1\Om_m\big)^{\!(1)}\big\}F^b 
     \Big]  \Big\}{\bigg|}_{t=0}\,.    
\end{align*}
 Let us now turn to equations~(\ref{SW-general-11}) for $N=2$.  Using 
 dimensional analysis and BRS covariance, it follows that the solution for
 $(\dl_2\Om_a)^{(2)}$ is an arbitrary linear combination
\begin{equation*}
   \big(\dl_2\Om_a\big)^{\!(2)}= 
        \sum_{i=1}^6 c_{i}\; \big(\dl_{2}\Om_a\big)^{(2)}_i
\end{equation*}
 with real coefficients $c_{i}$ of the linearly independent solutions
\begin{align*}
  \big(\dl_{2}\Om_a\big)^{(2)}_1&=\frac{\imu}{4}\>\th^{mn}\,\th^{rs}\,
      \big[\,D_a\Om_{mn} \,,\, \Om_{rs}\,\big]\\
  \big(\dl_{2}\Om_a\big)^{(2)}_2&=\frac{\imu}{4}\>\th^{mn}\,\th^{rs}\,
      \big[\,D_a \Om_{mr} \,,\, \Om_{ns}\,\big] \\
  \big(\dl_{2}\Om_a\big)^{(2)}_3&=\frac{\imu}{4}\>\th^{mn}\,\th^{rs}\;
     D_r\,\big[\,\Om_{mn}\,,\, \Om_{sa} \,\Big] \\
  \big(\dl_{2}\Om_a\big)^{(2)}_4&=\frac{1}{4}\>\th^{mn}\,\th^{rs}\;
      D_a\, \big(\,\hOm_{mn} \,\hOm_{rs} \big)\\
  \big(\dl_{2}\Om_a\big)^{(2)}_5&=\frac{1}{4}\>\th^{mn}\,\th^{rs}\;
      D_a\,\big(\, \Om_{mr} \, \Om_{ns}\big)  \\
  \big(\dl_{2}\Om_a\big)^{(2)}_6&=\frac{1}{4}\>\th^{mn}\,\th^{rs}\;
    D_r\,\big\{\Om_{mn} \,,\, \Om_{sa}\big\} \,.
\end{align*}
 Similarly, the solution for $\big(\dl_{2}F^a\big)^{(2)}$ is a linear 
 combination 
\begin{equation}
 \big(\dl_2 F^a)^{\!(2)} = \sum_{i=1}^3 p_{i}\; \big(\dl_{2} 
F^a\big)^{\!(2)}_i
          + \sum_{j=1}^{11} q_{j}\; \big(\dl_{2} F^a\big)^{\!(2)}_{j+3}
\label{Npsqs}
\end{equation}
 of linearly independent solutions
\begin{align*}
  \big(\dl_{2} F^a\big)^{(2)}_1 
     & = \frac{1}{4}\>\th^{mn}\,\th^{rs}\, \Om_{mn}\, \Om_{rs}\, F^a \\
  \big(\dl_{2} F^a\big)^{(2)}_2 
     & = \frac{1}{4}\>\th^{mn}\,\th^{rs}\, \Om_{mr}\,\Om_{ns} \, F^a  \\
  \big(\dl_{2} F^a\big)^{(2)}_3 & = \frac{\imu}{4}\>\th^{mn}\,\th^{rs}\, 
     \big(D_m\,\Om_{rs}\big) \,\big(\,D_n F^a \big) \\
  \big(\dl_{2} F^a\big)^{(2)}_4 & = \frac{1}{4}\>\th^{am}\,\th^{rs}\,
    \Om_{rs} \,\Om_{mb}\, F^b \\
  \big(\dl_{2} F^a\big)^{(2)}_5 & = \frac{1}{4}\>\th^{am}\,\th^{rs}\,
    \Om_{ms} \Om_{rb} \, F^b \\
  \big(\dl_{2} F^a\big)^{(2)}_6 & = \frac{1}{4}\>\th^{am}\,\th^{rs}\,
    \Om_{mb}\, \Om_{rs} \, F^b \\
  \big(\dl_{2} F^a\big)^{(2)}_7 & = \frac{1}{4}\>\th^{am}\,\th^{rs}\,
    \Om_{rb}\, \Om_{ms} \, F^b \\
  \big(\dl_{2} F^a\big)^{(2)}_8 & = \frac{\imu}{4}\>\th^{am}\,\th^{rs}\,
    \big(\,D_m \Om_{rs}\big)\, \big( D_b F^b \big)\\
  \big(\dl_{2} F^a\big)^{(2)}_9 & = \frac{\imu}{4}\>\th^{am}\,\th^{rs}\,
       \big(D_b\,\Om_{rs}\big)\,\big(D_m\,F^b\big) \\
  \big(\dl_{2} F^a\big)^{(2)}_{10} & = \frac{\imu}{4}\>\th^{am}\,\th^{rs}\,
       \big(D_b\,\Om_{ms}\big) \, \big(D_r\, F^b\big) \\
  \big(\dl_{2} F^a\big)^{(2)}_{11} & = \frac{\imu}{4}\>\th^{am}\,\th^{rs}\,
       \big( D_m\,\Om_{sb}\big) \,\big(D_r F^b\big) \\ 
  \big(\dl_{2} F^a\big)^{(2)}_{12} & = \frac{1}{4}\>\th^{am}\,\th^{rs}\,
       \Big( \big\{D_m , D_b\big\}\Om_{rs} \Big) \, F^b \\
  \big(\dl_{2} F^a\big)^{(2)}_{13} & = \frac{1}{4}\>\th^{am}\,\th^{rs}\,
       \Big( \big\{D_m , D_r\big\}\Om_{sb} \Big) \, F^b \\
  \big(\dl_{2} F^a\big)^{(2)}_{14} & = \frac{1}{4}\>\th^{am}\,\th^{rs}\,
       \Big( \big\{D_s , D_b\big\}\Om_{mr} \Big) \, F^b \,.
\end{align*}
 This is not a complete list of all independent solutions for 
 $(\dl_{2} F^a)^{(2)}$. For example, together with $(\dl_{2} F^a)^{(2)}_4$, one 
 also has the solution $\,\th^{am}\,\th^{rs}\, \Om_{rs}\,\{D_m\, D_b\}F^b$. 
 This, however, does not contribute to $\eL^{(2)}$ since, according to Step 1 
 in Subsection \ref{subsec:Five-One}, symmetrized products of more than one 
 covariant derivative $D_a$ acting on $F^a$ vanish. In the list above we have 
 omitted solutions with symmetrized products of Lorentz-covariant derivatives 
 acting on $F^b$.  By contrast, terms with symmetrized products$\{D_m,D_n\}$ 
 acting on $\Om_{ab}$ may give a non-vanishing contribution, since integration 
 by parts to move the covariant derivatives on $F^c$ will pick, upon 
 anti-symmetrization in $(m,n)$ a non vanishing contribution.  We finally note 
 that we have taken the coefficients $r,p_j,q_k$ to be real, to avoid 
 complexifications of the local $SO(1,3)$ symmetry into an $U(1,3)$ symmetry 
 and the difficulties that such complexifications, in terms of unwanted ghost 
 states, introduce~\cite{AG,Chams}.
 
 Since our interest here are corrections in $\th$ up to order two, it is enough
 to consider $\dl_1$ and $\dl_2$. Writing for the fields
\begin{equation}
   \begin{array}{c}
    \hOm_a'=\hOm_a + t\,\big(\dl_1\Om_a\big)^{\!(1)}  
     + t^2\,\big(\dl_1\Om_a\big)^{\!(2)} + t^2\,\big(\dl_2\Om_a\big)^{\!(2)} 
     + \ldots\\[5pt]
    \hF'^a =\hF^a + t\,\big(\dl_1F^a\big)^{\!(1)}  
    + t^2\,\big(\dl_1F^a\big)^{\!(2)} + t^2\,\big(\dl_2F^a\big)^{\!(2)} + \ldots
    \end{array}
\label{more-general}
\end{equation}
 we go over the construction in Sections 4 and 5. After quite a bit of work we
 obtain that there are no diffeomorphism invariant first order corrections to
 the Einstein--Hilbert Lagrangian, in agreement with the general arguments of
 Section \ref{sec:Two}. See Appendix C for intermediate results. For second order
 corrections we obtain that, while for \emph{pp}-wave metrics, second order
 contributions, for (2+2)-metrics there is a non-vanishing contribution, given 
 by
\begin{equation}
    \eL'^{(2)}_{\rm cov} =\frac{\sqrt{- h' \, h''}}{8\,\kappa^2}~\Big[ a\,I_1 
      - b\, \Big( R'+R''-\frac{\La}{2}\Big)\,I_5\Big]\,
\label{L2-cov-general}
\end{equation}
 where the coefficients $a$ and $b$ are given by eqs.~(\ref{a}) and (\ref{b})
 in terms of the coefficients $c,p,q,r$ and $c_i,p_j,q_k$ entering
 $(\dl_1\phi)^{(1)}$ and $(\dl_2\phi)^{(2)}$. Since $c,p,q,r$ and $c_i,p_j,q_k$
 are themselves arbitrary, the coefficients $a$ and $b$ are arbitrary. Putting
 together the Einstein--Hilbert action and its second order
 deformation~(\ref{L2-cov-general}), we reproduce the action written in
 eq.~(\ref{NC-action-general}).

\section{\label{sec:Seven}Conclusion and outlook}

 The Seiberg--Witten map can be viewed as a method to extend a local gauge
 symmetry to a larger symmetry living in the universal enveloping algebra of
 the original Lie algebra.  The method is not explicit, in the sense that it
 provides equations that must be solved for every gauge algebra. The solutions
 are power series in constant antisymmetric parameters $\th^{\m\n}$ whose
 coefficients depend polynomially on the fields involved and their derivatives.
 In the past, the solutions have been found to low orders in $\th$ for the
 gauge groups of particle physics, which in turn has led to
 anomaly-free~\cite{BMR} NC extensions of particle models~\cite{Mu}.

 In this paper we have solved the analogous problem for general relativity's
 symmetry group, namely the group of local Lorentz transformations.  This has
 resulted in a model for NC gravity whose classical action is a power series in
 a covariantly constant bivector $\,\th^{\m\n}(x)$. First and second order
 contributions to the classical action have been explicitely computed.  

 The condition $\na_\m\th^{\r\s}=0$ restricts four-dimensional geometries to
 \emph{pp}-wave metrics and direct sums of two two-dimensional metrics. For
 \emph{pp}-wave metrics, $\th^{\m\n}$ is null and first and second order
 corrections to general relativity's classical action vanish.  In turn,
 $(2+2)$-decomposable metrics correspond to either spacelike/timelike
 $\th^{\m\n}$. For them first order corrections vanish but second order
 corrections do not. The curvatures of the two two-dimensional metrics depend
 of $\th^{\m\n}$ and two arbitrary parameters $a$ and $b$, their
 $\th^{\m\n}\to 0$ limit being not smooth since $\th^{\m\n}$ cannot be zero.
 
 One of the motivations behind today's interest in NC gravity is studying
 whether non-commutativity may act as a source for gravity. From this point of
 view, the classical action obtained in this paper provides a field theory
 model with $\th^{\m\n}$ a ``gravity source''.  Furthermore, the family of
 $(2+2)$-geometries and their gravitational fields found here as can be
 understood as classically induced by non-commutativity through the
 Seiberg--Witten map.  Some solutions for induced, or emergent, NC gravity have
 been proposed within the context of matrix models~\cite{Steinacker}.

 In this paper we have not coupled gravity to matter. Matter couplings
 introduced in the classical action produce contributions of order one
 in $\th^{\m\n}$.  The simplest case is that of a $U(1)$ gauge field. One may
 keep gravity undeformed and only construct the Seiberg--Witten map for the 
 $U(1)$ field. Yet, by going to the generally covariant extension of the
 Seiberg--Witten construction along the lines explained here, the condition
 $\na_\r\th^{\m\n}=0$ comes in and one is again limited to \emph{pp}-wave 
 metrics and $(2+2)$-decomposable metrics. In this case~\cite{MR}, first order 
 contributions in the classical action provide NC deformations of 
 \emph{pp}-wave metrics.

 We want to finish with a few words about associativity of the Moyal--Groenewold
 product. It is precisely the fact that the deformation parameters $\th^{mn}$
 are constant in the Seiberg--Witten construction what ensures associativity.
 Turning the construction point into the origin of a locally inertial frame
 already destroys associativity. It is convenient to recall at this point that
 covariant constancy of $\th^{\m\n}$ does not ensure 
 associativity~\cite{Rivelles}. Furthermore, in the simple case of functions 
 on four-dimensional Euclidean space, the Moyal--Groenewold product for 
 bivectors of rank four, only is associative for constant 
 $\th^{\m\n}$~\cite{GGBRR}.

 It remains an open problem to extend the construction presented here to 
 metrics which are not of \emph{pp}-type or 2+2. For example, by considering 
 non-constant deformation parameters; or by solving the Seiberg--Witten 
 equations for other star products, e.g. Kontsevich's, for which the 
 deformation parameters form a Poisson tensor. However we are not aware of a 
 systematic way to compute Seiberg--Witten maps at higher order with
 such deformation parameters.

\begin{acknowledgments}
The Authors are grateful to MEC and UCM-CAM, Spain for partial support
through grants FIS2005-02309 and CCG07-UCM/ESP-2910.
\end{acknowledgments}

\appendix

\section{Solutions to equation~(\ref{master00})}

Being~(\ref{master00}) a quartic equation, its solutions can be determined
analytically. They can be cast in the form
\begin{equation*}
    \xi_i = y_i-\frac{p}{4}  \qquad i=1,2,3,4, 
\end{equation*}
 with $y_i$ given by
\begin{alignat*}{4}
  y_1 & = \frac{1}{2}\>\big(\sqrt{z_1} + \sqrt{z_2} + \sqrt{z_3}\big) & 
  y_3 & = \frac{1}{2}\>\big(- \sqrt{z_1} + \sqrt{z_2} - \sqrt{z_3}\big)
     \\[3pt] 
  y_2 & = \frac{1}{2}\>\big(\sqrt{z_1} - \sqrt{z_2} - \sqrt{z_3} \big) 
  \qquad & y_4 & = \frac{1}{2}\>\big(- \sqrt{z_1}  -  \sqrt{z_2} 
                       +  \sqrt{z_3}\big)
\end{alignat*}
 in terms of the solutions $z_1,z_2$ and $z_3$ of the resolvent cubic
 associated to equation~(\ref{master00}). These, in turn, have the following
 explicit expressions:
\begin{align*}
    z_1 & = \frac{p^2}{4} + s_1 + s_2 \\ 
    z_{\genfrac{}{}{0pt}{}{2}{3}}\! & = \frac{p^2}{4} 
       - \frac{1}{2}\> \big[\, (s_1+s_2) \mp
          \imu\,\sqrt{3}\,(s_1-s_2)\,  \big]  
\end{align*}
 where $s_1$ and $s_2$ are the cubic roots
\begin{equation*}
    s_{\genfrac{}{}{0pt}{}{1}{2}}\! = {\bigg[\, \frac{p^2}{2k^2}\>(k+1) 
           \pm \frac{p^2}{2k^2}\> \sqrt{\Dl(k,p^2)}\, \bigg]}^{1/3}\,,
\end{equation*}
 $\Dl(k,p^2)$ is the discriminant 
\begin{equation*}
    \Dl(k,p^2) 
    = (k+1)^2 + \frac{4\,(p^2 -4)^3k}{27p^4} 
    = \big[k-k_+(p^2)\big]\,\big[k-k_-(p^2)\big] 
\end{equation*}
 and $k_\pm(p^2)$ are as in~(\ref{kpm}).The real or complex nature of
 $z_1,z_2,z_3$, hence of the solutions $\xi_i$, depend on the sign of $\Dl$. To
 study $\Dl$, we note that
\begin{alignat*}{6}
   & k_-\!\geq k_+\!> 0 &\quad & \textnormal{for}& \quad  & p^2\!\leq 1 \\
   & k_-\!=k_+^* &   & \textnormal{for}&   & 1<p^2\!<4\\
   & k_-\!\leq k_+\!<0 &   & \textnormal{for}&   & 4\leq p^2\,. 
\end{alignat*}
 Attending to the sign of $\Dl$, the $(p^2,k)$-domain can be divided into the 
 following subdomains:
\begin{align*}
  & D_>\! = \{p^2\!\leq 1,\,k>0\} \,\cup\,  \{p^2\!<1,\,k>k_- \} 
          \,\cup\, \{p^2<1,\, k<k_+\} \\
  & D_0 =  \{p^2\!<1,\, k=k_-\} \,\cup\, \{p^2\!<1,\, k=k_+\} \\
  & D_<\! = \{p^2\!<1,\, k_+\!<k<k_-\}
\end{align*}
 We now have
\begin{itemize}
\vspace{-9pt}
\item[(1)] In $D_>$, the discriminant $\Dl$ is positive, $z_1$ is real and
 positive and $z_2,z_3$ are complex conjugate of each other.  This implies
 that there are two real and two complex solutions $\xi_i$. 
\vspace{-7.5pt}
\item[(2)] In $D_0$, $\Dl$ vanishes and $z_1$
 and $z_2\!=z_3$ are real and positive. Hence there are four real solutions
 $\xi_i$, with at most three of them distinct.
 \vspace{-7.5pt}
\item[(3)] Finally, in $D_<$, $\Dl$ is negative and one can easily 
 deduce that
\begin{equation}
   k\,<\, \frac{36}{k}\,{\bigg( \frac{k+1}{4-p^2}\bigg)}^2 \!
      <\, \frac{16\,(4-p^2)}{3\,p^4}\,.
\label{beta}
\end{equation}
 We are only interested in positive $\,z_2$ and $z_3$, since otherwise the four
 $y_i$ are complex and there are no real solutions for $\xi$. From Vieta's 
 relation we obtain
\begin{equation*}
 z_1\,z_2 + z_2\,z_3 + z_3\,z_1 = \frac{3\,p^4}{16} - \frac{4-p^2}{k}\!>\,0 \,,
\label{gamma}
\end{equation*}
 in contradiction with~(\ref{beta}). We conclude that in $D_<$ there are no
 real solutions for $\xi$, thus proving the restriction~(\ref{restriction}).
\vspace{-9pt}
\end{itemize}

\section{The vierbein postulate and BRS Lorentz symmetry}

We collect here some technical issues concerning Subsection
\ref{subsec:Five-One}. First we prove the equivalence between
equations~(\ref{VP-bz}), (\ref{torsion-free}) and~(\ref{SVP}), (\ref{STF}).
Going from~(\ref{VP-bz}), (\ref{torsion-free}) to~(\ref{SVP}), (\ref{STF}) is
straightforward. Indeed, condition $\Ga^b_{cd}(x)=0$ in~(\ref{STF}) holds
trivially, for $x^a$ is the origin of a locally inertial frame. Furthermore,
since~(\ref{VP-bz}) and (\ref{torsion-free}) hold for arbitrary $\bx^a$ close
to $x^a$, one may act on them with products of $\bar{\pa}_a$ and
$\,\bar{\na}_a - \imu \,\bar{\Om}_a$ and then set $\bar{x}^a = x^a$. This
leads to~(\ref{SVP}), (\ref{STF}).  Let us now prove that~(\ref{SVP}),
(\ref{STF}) imply~(\ref{VP-bz}), (\ref{torsion-free}).  Expanding
$\,(\bar{\na}_a - \imu\,\bar{\Om}_a) \bar{F}^c$ in power series of
$\,\bx^a-x^a\!$, we have
\begin{equation}
  \left( \bar{\na}_b - \imu \,\bar{\Om}_b \right) \bar{F}^c
      = \sum_{k=1}^\infty \, \frac{1}{k!} \,  (\bx-x)^{a_1} \,
         \cdots \, (\bx-x)^{a_k} \, \pa_{a_1} \cdots \pa_{a_k}
         \left(\na_b - \imu\,\Om_b\right) F^c (x)\, .
\label{IVP-expand}
\end{equation}
 For the first term in the sum, it follows from eq.~(\ref{SVP}) that
\begin{equation}
   \pa_{a_k} \left(\na_b -\imu\,\Om_b\right) F^c(x) =
      \imu \, \Om_{a_k}(x) \, \left(\na_b -\imu\,\Om_b\right) F^c (x)
       = 0 \, .
\label{induction}
\end{equation}
 Acting with $\pa_{a_k-1}$ on~(\ref{induction}), we obtain for the second term
 in the sum~(\ref{IVP-expand})
\begin{equation*}
  \pa_{a_{k-1}} \pa_{a_k} \left(\na_b -\imu\,\Om_b\right) F^c(x) 
     = \imu\,\big[\pa_{a_{k-1}}\Om_k\big]\,\big[(\na_b-\imu\,\Om_b)F^c(x)\big]
     + \Om_{a_k}(x)\, 
       \big[\pa_{a_{k-1}} \big(\na_b -\imu\,\Om_b\big) F^c (x)\big] \,.
\end{equation*}
 The first contribution on the right-hand side vanishes because of~(\ref{SVP}), 
 the second one because of~(\ref{induction}). Repeating the argument we arrive 
 at
\begin{equation*}
   \pa_{a_1} \pa_{a_2} \cdots \pa_{a_k} \,
     \left(\na_b - \imu\,\Om_b \right) F^c(x) = 0 \, ,
\end{equation*}
 which implies~(\ref{VP-bz}) upon substitution in~(\ref{IVP-expand}). Similarly, 
 expanding $\Ga_{cd}^b (\bx)$ about~$x^a$ and using eq.~(\ref{STF}), we have
\begin{equation*}
   \Ga_{cd}^b (\bx)
     = \sum_{k=0}^\infty \, \frac{1}{k!} \,
        (\bx-x)^{a_1} \, \cdots \, (\bx-x )^{a_k} \,
        \pa_{a_1} \cdots \pa_{a_k} \Ga_{cd}^b (x) = \Ga_{dc}^b(\bx) \, ,
\end{equation*}
 which is the torsion-free condition~(\ref{torsion-free}). 

 To illustrate this equivalence, let us show that the solutions to~(\ref{SVP}), 
 (\ref{STF}) can be retrieved from those to~(\ref{VP-bz}), (\ref{torsion-free}) 
 by taking derivatives $\bar{\pa}_{a_1}\cdots\bar{\pa}_{a_n}$ and setting 
 $\bx^a=x^a$.  Indeed, the solutions to ~(\ref{VP-bz}), (\ref{torsion-free}) at 
 any point $\bx^a$ of locally inertial frame with origin $x^a$ is known to be 
\begin{equation}  
   \bar{\Om}_a = \frac{\imu}{2} \, \left\{  \bar{g}_{ab}\, \big[\,  
       \bar{F}^c \big( \bar{\pa}_c  \bar{E}^b\big) 
     - \big(  \bar{\pa}_c  \bar{F}^b \big)  \bar{E}^c \,\big] 
    +  \bar{\cE}_b \,  \bar{\pa}_a  \bar{E}^b 
    + \left(  \bar{\pa}_b  \bar{\cE}_a \right)  \bar{E}^b 
    - \big(  \bar{\pa}_a \bar{F}^b \big)  \bar{\cF}_b 
    -  \bar{F}^b \,  \bar{\pa}_b  \bar{\cF}_a \right\} \;
\label{solution-Omega}
\end{equation}
 for the spin-connection, and
\begin{equation}
     \bar{\Ga}_{cd}^b = - \frac{1}{2} \, \Big[ 
         \bar{g}^{ba}\, \big( \bar{\pa}_a \, \bar{g}_{cd}\big) 
       + \bar{g}_{ac} \,\big(  \bar{\pa}_d\,  \bar{g}^{ba}\big) 
       + \bar{g}_{ad}\, \big(  \bar{\pa}_c\,  \bar{g}^{ba}\big) \Big]  
\label{solution-Christoffel}
\end{equation}
 for the Christoffel symbols, with 
$\,\bar{g}_{ab}=\bar{\cF}_a \bar{\cE}_b = \bar{\cF}_b\, \bar{\cE}_a\,$ the 
metric and $\,\bar{g}^{ab} = \bar{E}^a\,\bar{F}^b=\bar{E}^b\,\bar{F}^a\,$ the 
inverse metric. By taking $\bx^a = x^a$ in (\ref{solution-Omega}) and by using 
that at the origin of a locally inertial frame
\begin{alignat*}{5}
   & {\pa}_c\, {g}^{ab} =0 & ~\Rightarrow~ 
     & \big( {\pa}_c\,{E}^a\big) {F}^b 
         + {E}^a  \big( {\pa}_c\,{F}^b\big) =0 
       &~\Rightarrow~ & {\cE}_a\,\big({\pa}_c\,{F}^a\big) 
                        + \big({\pa}_c\, {F}^b\big) {\cF}_b =0 \\
  & {\pa}_c\, {g}_{ab} =0 & ~\Rightarrow~ 
     &  ~~\big( {\pa}_c{\cF}_a \big){\cE}_b 
         + {\cF}_a \big( {\pa}_c{\cE}_b\big) =0 
       &~\Rightarrow~ & ~~ {g}_{ab}\big({\pa}_c\,{E}^b\big) 
                         - {\pa}_c{\cF}_a =0\,, 
\end{alignat*}
 we obtain
\begin{equation*}
     \Om_a = -\imu \,\big(\pa_a F^b\big)\, \cF_b\,.
\end{equation*}
 This is precisely the solution to equation~(\ref{STF}) for $n=1$.

 Next we show the consistency of the BRS operator on which the Seiberg--Witten
 construction for the Lagrangian is based with the vierbein
 postulate~(\ref{VP-bz}) and the torsion-free condition~(\ref{torsion-free}).
 This amounts to proving that $\,s\bar{\Ga}^b_{cd}=0$ and that $\bar{\Om}_a$
 in~(\ref{solution-Omega}) transforms as in~(\ref{C-BRS}). From
 eqs.~(\ref{inverse-F}), (\ref{inverse-E}) and the transformation law
 $\,s\bar{F}^a=\imu\,\bar{\la} \bar{F}^a$, it follows that
\begin{equation}
  s\bar{E}^a=-\imu\,\bar{E}_a\,\bar{\la}   \qquad     
  s\bar{\cE}_a=\imu \,\bar{\la}\,\bar{\cE}_a    \qquad       
  s\bar{\cF}_a = -\imu\,\,\bar{\cF}_a\,\bar{\la}\,.
\label{other-sL}
\end{equation} 
 Eqs.~(\ref{other-sL}), $\bar{g}_{ab}=\bar{\cF}_a \bar{\cE}_b$ and
 $\bar{g}^{ab}=\bar{E}^a\,\bar{F}^b$ imply 
 $\,s\bar{g}_{ab}= s(\pa_c\bar{g}_{ab})=0$ and 
 $s\bar{g}^{ab}= s(\bar{\pa}_c\,\bar{g}^{ab})=0$. Upon substitution 
 in~(\ref{solution-Christoffel}) one has $\,s\bar{\Ga}^b_{cd}=0$. Analogously, 
 acting with $s$ on eq.~(\ref{solution-Omega}), employing~(\ref{other-sL}) and 
 simplifying with the help of~(\ref{inverse-F}) and (\ref{inverse-E}), we obtain 
 after some simple algebra that 
 $s\bar{\Om}_a= \bar{\pa}_a\bar{\la}-\imu\,[\bar{\Om}_a,\bar{\la}]$, in 
 agreement with~(\ref{C-BRS}).

\section{Derivation of eq.~(\ref{L2-cov-general})}

 Eq.~(\ref{L2-general}) is obtained by using the construction explained in
 sections \ref{sec:Four} and \ref{sec:Five} to the Seiberg--Witten maps 
 $\hOm'_a$ and $\hF'^a$ in eqs.~(\ref{more-general}). In this Appendix we 
 present some partial results of this computation. 
 Equation~(\ref{L-corrections}) gives for the first order contribution
\begin{equation*}
   \eL^{\prime(1)} = \eL^{(1)} + \dl\eL^{(1)}\,,
\end{equation*}
 where $\eL^{(1)}$ is as in eq.~(\ref{cL1}) and $\dl\eL^{(1)}$ reads 
\begin{equation*}
   \dl\eL^{(1)} =   \frac{e}{2\kappa^2}~\th^{mn}\,\Big[\,
     (c-p)\,  E^a\,R_{ab}\,R_{mnc}{}^b F^c 
     + q\,R_{mn} \big(R_{ab}\,E^a\,F^b -\La\big) \,\Big]\,.
\end{equation*}
 The coefficients $c,\,p$ and $q$ are those in $\dl_1\hOm_a$ and $\dl_1\hF^a$.
 From Section \ref{sec:Five} we know that $\eL^{(1)}_{\rm cov}=0$. For the 
 contribution $\dl\eL^{(1)}$, Steps 1 and 2 in Subsection \ref{subsec:Five-One} 
 yield $\dl\eL^{(1)}_{\rm cov}=0$. Hence $\eL'^{(1)}_{\rm cov}=0$.

 The second order contribution can also be written as
\begin{equation*}
    \tilde{\eL}^{(2)} = \eL^{(2)} + \dl\eL^{(2)}
\end{equation*}
 where $\,\eL^{(2)}\,$ is as in eq.~(\ref{cL2}) and $\,\dl\eL^{(2)}\,$ has a
 very complicated expression. Here we only display its result after going
 through the `covariantization procedure' of Steps 1 and 2 in Section 
 \ref{sec:Five}:
\begin{align}
    \dl\eL^{(2)}_{\rm cov} = \frac{\sqrt{- g}}{16\kappa^2}~& \Big[\, m_6\,
   {\cal R}_6 
     + m_7\,{\cal R}_7 + m_8\, {\cal R}_8 +m_9\, {\cal R}_{9} 
     + m_{10}\, {\cal R}_{10} + m_{11}\, {\cal R}_{11} \notag  \\[2pt]
   & \! + \big( R-\La\big)\,\big( - n_1 {\cal Q}_1 + n_2 {\cal Q}_2 
           +  n_3 {\cal Q}_3 - q^2 {\cal Q}_4 \big) \notag \\
   & \! + 2\,(2c_1+c^2)\,\big(4\,{\cal B}_3-{\cal B}_4\big)
        + 4\,c_2 \,\big(4\,{\cal B}_5-{\cal B}_6\big) 
        + 2\,c_3\, \big(2\,{\cal B}_7+{\cal B}_8\big)\Big] \,, 
   \label{L2-general}
\end{align}
 The coefficients $m_6,\cdots,m_{11}$ and $n_1,n_2,n_3$ are given in terms of 
 those in $\dl_1\hat{\phi}$ and $\dl_2\hat{\phi}$ by
\begin{align*}
    m_6 & = - 2r- 2q^2 + 4q_2 - q_7 \\
    m_7 & = (c - p)^2 \\
    m_8 & = 4pq - 4q_1 + (q_5-q_6) \\
    m_9 & = - r + 2q_4 + \frac{1}{2}\>q_8 \\
    m_{10} & = 2q_3 \\
    m_{11} & = 4\,(pc-p^2-c^2+p_1- p_3 + c_1) + 2\,(p_2 +c_2)
\end{align*}
 and
\begin{align*}
    n_1 &= \frac{r}{2} + 2p_2+q_4 \\
    n_2 & = p^2 -pq - 2p_1 +q_3 \\
    n_3 &= \frac{r}{2} + q^2 +2q_1-q_2\,.
\end {align*}
 With respect to $\,\eL^{(2)}_{\rm cov}$, new invariants occur in
 $\,\dl\eL^{(2)}_{\rm cov}$, namely
\begin{equation*}
   {\cal R}_{11} =  \th^{\m\n}\,\th^{\r\s}\, g^{\ga\dl}\, 
       R_{\a\b}\,R_{\m\n\ga}{\!}^\a\,R_{\r\s\dl}{\!}^\b 
\end{equation*}
 and
\begin{align*}
   {\cal Q}_2 & = \th^{\m\n}\,\th^{\r\s}\, 
      R_{\m\n\a\b}\,R_{\r\s}{\!}^{\a\b} \\
   {\cal Q}_3 & = \th^{\m\n}\,\th^{\r\s}\, R_{\m\n\r}{\!}^\a\, R_{\s\a} \\
   {\cal Q}_4 & = \th^{\m\n}\,\th^{\r\s}\, R_{\m\r}\,R_{\n\s} \\
   {\cal B}_3 & = \th^{\m\n}\,\th^{\r\s}\, \na_\a \big( 
         R_{\m\n}{\!}^{\a\b}\,\na_\r R_{\s\b} \big) \\
   {\cal B}_4 & = \th^{\m\n}\,\th^{\r\s}\, \na_\a \na_\b \big( 
         R_{\m\n\ga}{\!}^{\a}\,R_{\r\s}{\!}^{\ga\b}\big) \\
   {\cal B}_5 & = \th^{\m\n}\,\th^{\r\s}\, \na_\a \big( 
         R_{\m\r}{\!}^{\a\b}\,\na_\n R_{\s\b} \big) \\
   {\cal B}_6 & = \th^{\m\n}\,\th^{\r\s}\, \na_\a \na_\b \big( 
         R_{\m\r\ga}{\!}^{\a}\,R_{\n\s}{\!}^{\ga\b}\big)  \\
   {\cal B}_7 & = \th^{\m\n}\,\th^{\r\s}\, \na_\a  \na_\m \big( 
         R_{\r\s}{\!}^{\a\ga}\,R_{\n\ga}\big) \\
   {\cal B}_8 & = \th^{\m\n}\,\th^{\r\s}\, \na_\a \na_\m \big(
         R_{\r\s}{\!}^{\ga\dl}\,R_{\ga\dl\n}{\!}^{\a}\big)\,.
\end{align*}
 We only have to compute these invariants for \emph{pp}-wave metrics and
 $(2+2)$-decomposable metrics. For \emph{pp}-wave metrics, they vanish
 identically, in agreement with the discussion of Section \ref{sec:Two}. For
 $(2+2)$-decomposable metrics, they become 
\begin{align*}
     {\cal R}_{11} & = -8I_1  \\
     {\cal Q}_2 & = 2\,{\cal Q}_3 = 4\,{\cal Q}_4  = -8\,I_5 \\
     4\,{\cal B}_3 & =  {\cal B}_4 =  8\,{\cal B}_5  
     = 2\,{\cal B}_6  =  2\,{\cal B}_7  = - {\cal B}_8 = - 4\,J_1\,, 
\end{align*}
 with $I_1,\,I_5$ and $J_1$ as in eqs.~(\ref{I1}),~(\ref{I5}) and~(\ref{J1}).
 Substituting in (\ref{L2-general}) and summing the
 contribution~(\ref{L2-particular-22}) from $\eL^{(2)}$, we reproduce
 eq.~(\ref{NC-action-general}), with the coefficients $a$ and $b$ given by
\begin{align}
   a & = \frac{1}{2} - \frac{1}{2} \, m_6 - 4 \,m_7 - m_8 + m_9 + 2 \, m_{10} 
- 2 \, m_{11} \label{a}\\
   b & = \frac{1}{2} - 4n_1 + 8n_2 + 4n_3 - q^2\,.  \label{b}
\end{align}

\end{document}